\documentclass[12pt,preprint]{aastex63}   

\shorttitle{CO Maps of four GMCs}
\shortauthors{Bieging \& Kong}

\newcommand{\kms}{km~s$^{-1}$}

\newcommand{\msun}{M$_{\odot}$}

\newcommand{\tco}{$^{13}$CO}        

\def\av#1{\langle#1\rangle}	      

\begin{document}


\title{
The Star Formation---Gas Density Relation in Four Galactic GMCs:\\Effects of Stellar Feedback}

\author[0000-0002-6291-7805]{John H. Bieging}
\affil{Steward Observatory, The University of Arizona, Tucson, AZ 85721, USA}

\author[0000-0002-8469-2029]{Shuo Kong}
\affil{Steward Observatory, The University of Arizona, Tucson, AZ 85721, USA}

\accepted{for The Astrophysical Journal}



\begin{abstract}
We present maps of 4 galactic giant molecular clouds (GMCs) in the J=2-1 emission of both CO and \tco.  We use an LTE analysis to derive maps of the CO excitation temperature and column density and the distribution of total molecular gas column density, $\Sigma_{gas}$. The depletion of CO by freeze-out onto cold dust grains is accounted for by an approximation to the results of \citet{Lewis_2021} which were derived from far-IR observations with {\it Herschel}.  The surface density of young stellar objects (YSOs) is obtained from published catalogs.  The mean YSO surface density exhibits a power-law dependence on $\Sigma_{gas}$, with exponents in the range 0.9 to 1.9.  Gas column density probability distribution functions (PDFs) show power-law tails extending to high column densities. The distributions of sonic Mach number, $M_S$ are sharply peaked at $M_S \sim 5 - 8$ for 3 GMCs; a fourth has a broad distribution up to $M_S =30$, possibly a result of feedback effects from multiple OB stars.  An analysis following the methodology of \citet{2021ApJ...912L..19P} finds that our sample of GMCs shows power-law relations that are somewhat shallower than found by \citet{2021ApJ...912L..19P} for the star formation rate vs. $\av{\Sigma_{gas}}$ and vs. $\av{\Sigma_{gas}}/t_{ff}$ in a different sample of clouds.  We discuss possible differences in the two samples of star-forming clouds and the effects of stellar feedback on the relation between gas density and star formation rate.

\end{abstract}






\section{Introduction}
This paper presents new results on the distribution and properties of molecular gas in 4 Galactic giant molecular clouds (GMCs) that have been studied extensively at many wavelengths and in many tracers of their gas, dust, and stellar content.  We examine the relation between the molecular gas column density and the current rate of star formation, as well as the area distribution of the GMCs as a function of the gas column density.  Together these (empirical) relations should determine the overall rate of star formation within the molecular clouds.  These observations also provide constraints on theoretical models for star formation in GMCs from physical scales of  embedded stellar clusters ($< 1$~pc) up to entire GMCs ($\sim 50$~pc).  Currently favored models employ so-called gravo-turbulent mechanisms to compress the gas into star-forming clumps via dissipation of supersonic turbulence followed by gravitational collapse \citep[e.g.,][]{1994ApJ...423..681V,2012A&ARv..20...55H}.

On much larger scales ranging from kiloparsecs to entire galaxies, in his pioneering work \citet{1959ApJ...129..243S,1963ApJ...137..758S} found that the rate of star formation in the Milky Way was a power-law function of the mean interstellar gas density, and considering a variety of observational constraints, with a most likely exponent of 2.  Subsequent work led by Kennicutt and many others refined this conclusion, finding that a power-law exponent of $\sim1.4$~ applies for most galaxy types over 5 orders of magnitude in gas {\it surface} density \citep[see the review by][]{2012ARA&A..50..531K}.  In the present study we find that these galactic scale relations appear to hold even down to the sub-parsec scale of star-forming molecular clumps.

The present study is a continuation of a series of papers reporting our mapping observations of CO isotopologues in the J=2-1 and J=3-2 rotational transitions toward a selection of Galactic molecular clouds.  Regions previously reported include the Sh2-254---258 group of H~II regions \citep{2009AJ....138..975B}; W51 \citep{2010ApJS..191..232B}; W3 \citep{2011ApJS..196...18B}; Serpens Main \citep{2013ApJS..209...39B}; NGC~1333 \citep{2014ApJS..214....7B}; Sh2-235 \citep{2016ApJS..226...13B}; and Cep B and C, adjacent to the Cep OB3 association \citep{2018ApJS..238...20B}.  The last two studies compared maps of the molecular gas column density derived from CO observations, with the distribution of young stellar objects (YSOs);  in both cases we found that the surface density of YSOs showed a power-law dependence on the molecular gas column density.  If one assumes both a mean stellar mass and a mean age for the cataloged YSOs, the derived relations could be expressed as a rate of star formation vs. molecular gas column density.  

In the case of Sh2-235, our analysis made use of a previous census by \citet{2014MNRAS.439.3719C} of YSOs in the embedded stellar clusters forming from the molecular gas.  We obtained fully sampled maps of CO isotopologues in the J=2-1 and 3-2 rotational transitions.  Comparison of these maps indicated that for the bulk of the molecular gas, the CO rotational levels were populated very close to LTE conditions.  With this evidence as justification, we derived the molecular gas column density by a standard LTE analysis and compared the gas column density with the surface density of YSOs \citep{2016ApJS..226...13B}.  The inferred relationship was well described by a power law in molecular gas column density, with a power-law exponent of 1.63.  The area distribution of gas column density (the {\it structure function} as defined below) was an exponential function of the column density, in agreement with other studies that employed different observational methods.  

The molecular clouds associated with Cep OB3 also showed a power-law dependence of YSO surface density on molecular gas column density \citep{2018ApJS..238...20B}.  The number of YSOs was smaller than for Sh2-235, however,  so the statistical errors were somewhat larger for Cep OB3.  We found a power-law exponent of 1.74, a little steeper than for Sh2-235.  

Previous studies of YSO surface density as a function of gas column density have mainly considered only GMCs closer than 1 kpc.  Since GMCs are recognized as the dominant star-forming entities in the Galaxy, it is important to extend such studies to a larger sample of GMCs, which inevitably means those at distances greater than 1 kpc from the Sun.  A meaningful result, however,  requires catalogs of YSOs which reach fainter apparent magnitudes than the 2MASS survey achieved.  The paper by \citet{2014MNRAS.439.3719C}  presented catalogs of YSOs for four other Galactic GMCs in addition to Sh2-235, all with comparable completeness.   Their census of YSOs was made from JHK$_s$ images from the Flamingos camera on the Kitt Peak 2.1 m telescope, and also from the SAO Wide Field Camera on the 6.5 m Multiple Mirror Telescope on Mt. Hopkins, Arizona.  Their NIR photometric data were significantly deeper than the 2MASS all sky survey.  Additional observations from the {\it Spitzer}  Gould's Belt Project were used to classify the stellar sources into proto-stellar class I and II categories, as well as objects clearly identifiable as YSOs but of uncertain evolutionary status.  The four other GMCs besides Sh2-235 in the \citet{2014MNRAS.439.3719C} paper included W5-east, Sh2-252, Sh2-254---258, and NGC~7538.  Our previous work on Sh2-235 \citep{2016ApJS..226...13B} motivated the present study, in which we show maps of the molecular gas associated with these four GMCs, and examine the correlation between the distribution of YSOs determined by \citet{2014MNRAS.439.3719C} with that of the molecular gas determined from our own observations.  We also use the kinematical information from the CO spectra to examine supersonic turbulence in the gas, as a constraint on theoretical gravo-turbulent models of star formation.  


The four GMCs in this study are all currently forming stellar clusters, some of which contain one or more OB stars with associated H~II regions (as indicated by the names mainly drawn from previous catalogs of Galactic nebulae containing ionized gas, e.g., \citealt{1959ApJS....4..257S} and \citealt{1958BAN....14..215W}).  The presence of newly formed OB stars will of course have a significant effect on the physical conditions in the adjacent molecular gas, thereby influencing the ongoing process of star formation in the cloud.  We will consider such effects in the discussion later in this paper.

In this work, we adopt the distances used by \citet{2014MNRAS.439.3719C} to be consistent with their catalog of YSOs and related quantities.  (We examined the {\it Gaia} EDR3 distances for the YSOs included in that catalog.  The spread of values for each of the GMCs is relatively large but consistent with our adopted distances.) All four GMCs lie in the range 1.6 to 2.7 kpc from the Sun.  Table \ref{tab:gmc} lists the adopted distances and the location and dimensions of the mapped fields.  

The published literature on observations for these 4 GMCs is voluminous.  \citet{2014MNRAS.439.3719C} give concise summaries of that work for W5-east, Sh2-252, Sh2-235, and NGC~7538, emphasizing the young stellar clusters, H~II regions and their exciting stars, and compact IR sources revealed by previous observations.  A similar discussion for Sh2-254---258 can be found in \citet{2008ApJ...682..445C} and also in \citet{2009AJ....138..975B} which presents our earlier CO mapping study of this H~II region complex.  

\section{Observations and Data Reduction}

All of the observations presented here were made with the Heinrich Hertz Submillimeter Telescope\footnote{For technical specifications, see the website \url{https://aro.as.arizona.edu/?q=facilities/submillimeter-telescope}.} on Mt. Graham, Arizona, at an elevation of 3200 m.  This facility is operated by the Arizona Radio Observatories, a division of Steward Observatory at the University of Arizona.

Procedures for the observations of the J=2-1 transitions of $^{12}$C$^{16}$O (hereafter CO) and $^{13}$C$^{16}$O (hereafter \tco) were identical to those described in \citet{2014ApJS..214....7B}.  The receiver was the dual-polarization ALMA Band 6 prototype sideband separating mixer system (Lauria et al. 2006).  The spectrometer was a set of four filterbanks each with 512 channels of 1.0 MHz bandwidth and separation (corresponding to a velocity resolution of $\sim$1.3 \kms~for the J=2-1 lines).  Spectra for the vertical and horizontal polarizations and the upper and lower mixer sidebands were processed independently.  The CO line at $\sim$230 GHz is detected in the mixer upper sideband and the \tco~ line at $\sim$220 GHz in the lower sideband.  Each line is observed in both of the orthogonal linear polarizations, providing independent maps of the emission which are averaged together to reduce the noise by $\sim 1/\sqrt{2}$.  Observations were made during the period 2019 December 11 through 2020 March 2.  The fields to be mapped were divided into contiguous $10\arcmin \times 10\arcmin$ ``tiles", each of which was observed in the standard On-The-Fly (OTF) scanning mode.  The data were calibrated and processed as described in \citet{2014ApJS..214....7B}.  The line intensity calibrators were the compact molecular source W3(OH) and the central CO line peak of W51D, with main-beam brightness temperatures and integrated line intensities as given in \citet{2011ApJS..196...18B}.  Telescope pointing was checked and corrected as necessary approximately every 2 hours.  The typical pointing error was $< 5\arcsec$.  Since the transitions of CO and \tco~are observed simultaneously through identical telescope optics and separated in the sidebands of the mixer (see Lauria et al. 2006), the registration of the two isotopologue maps is guaranteed to be correct, ensuring the fidelity of the LTE analysis (discussed below) that derives gas kinetic temperature from CO and molecular column density from \tco.  Pointing errors were in any case always a small fraction of the diffraction-limited beamwidth (FWHM $\approx 32\arcsec$).  

We applied spatial smoothing to the J=2-1 maps, as with other papers in this series \citep[see e.g.,][]{2013ApJS..209...39B} by convolution of the maps with Gaussian kernels.  This smoothing ensures that the CO and \tco~ maps have the same resolution (since the diffraction limited beam sizes differ by the ratio of the line frequencies, $\sim$ 5\%).  Spatial smoothing also reduces the noise per pixel by roughly 40\%.  The effective resolution after convolution is 38\arcsec~(FWHM) for both isotopologues.  All of the J=2-1 data presented here have been smoothed to this resolution.  Typical rms noise values per pixel in each 1 MHz filter channel calculated over emission free velocity ranges, were 0.09 K in beam-smoothed brightness temperature for CO and 0.10 K for \tco~ for all 4 mapped fields.

Since the line rest frequencies differ by $\sim5\%$, the velocity sampling is not the same for the two isotopologues.  Therefore, we resampled the maps by 3rd-order interpolation in velocity on identical LSR values at 0.45 \kms~intervals, i.e., slightly better than Nyquist sampling. Our LTE analysis employs only velocity moments of the data cubes as well as peak line brightness temperatures, so the velocity re-sampling ensures that the moments are calculated over precisely the same range in velocity.

\section{Results}
\subsection{Global spectra}
\begin{figure}
    \plotone{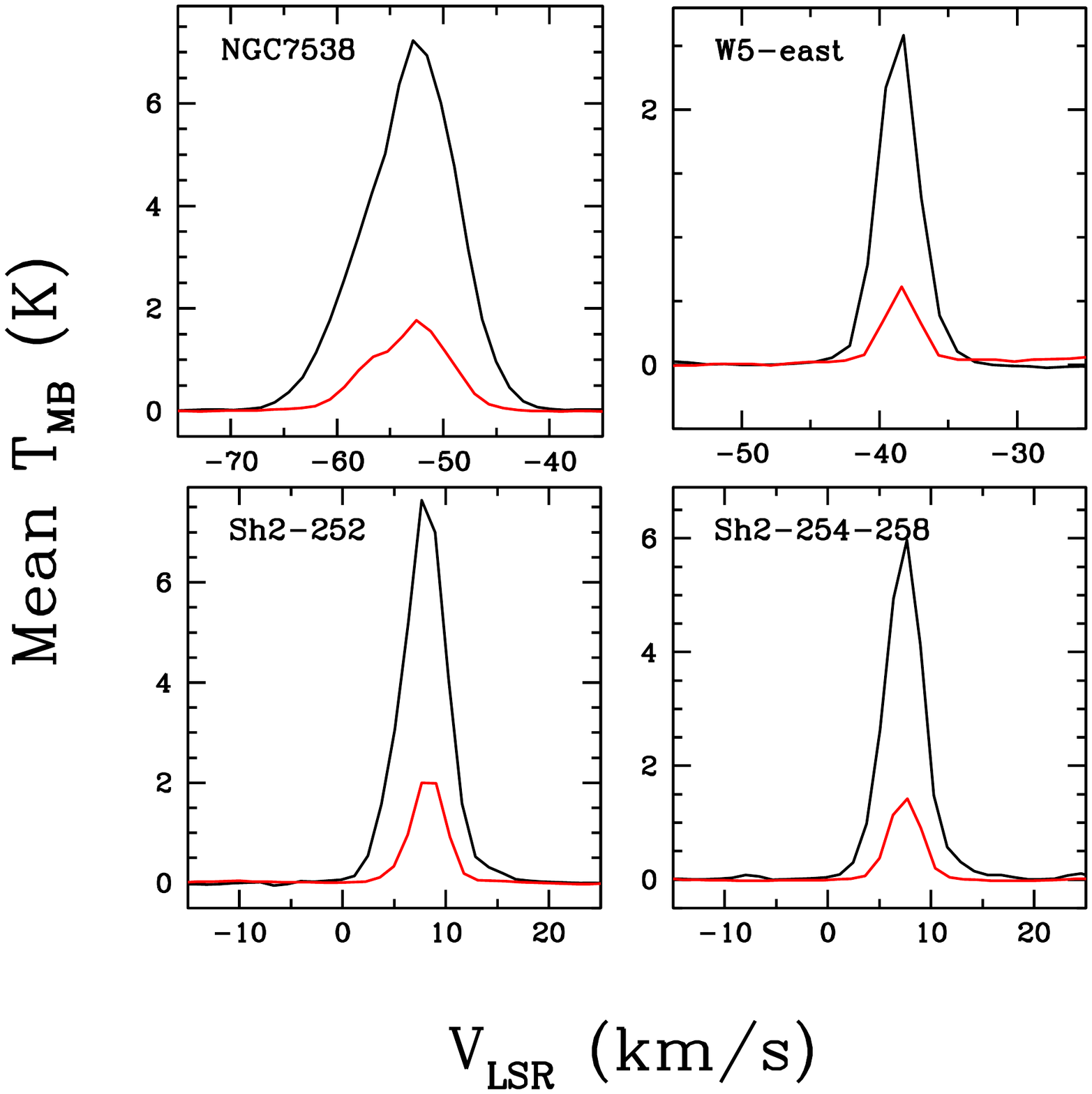}
    \caption{Main beam brightness temperature spectra for CO (black) and \tco ~J=2-1 (red) transitions, of all 4 GMCs, averaged over each mapped field.}\label{fig:1}
\end{figure}

\begin{figure}
\plotone{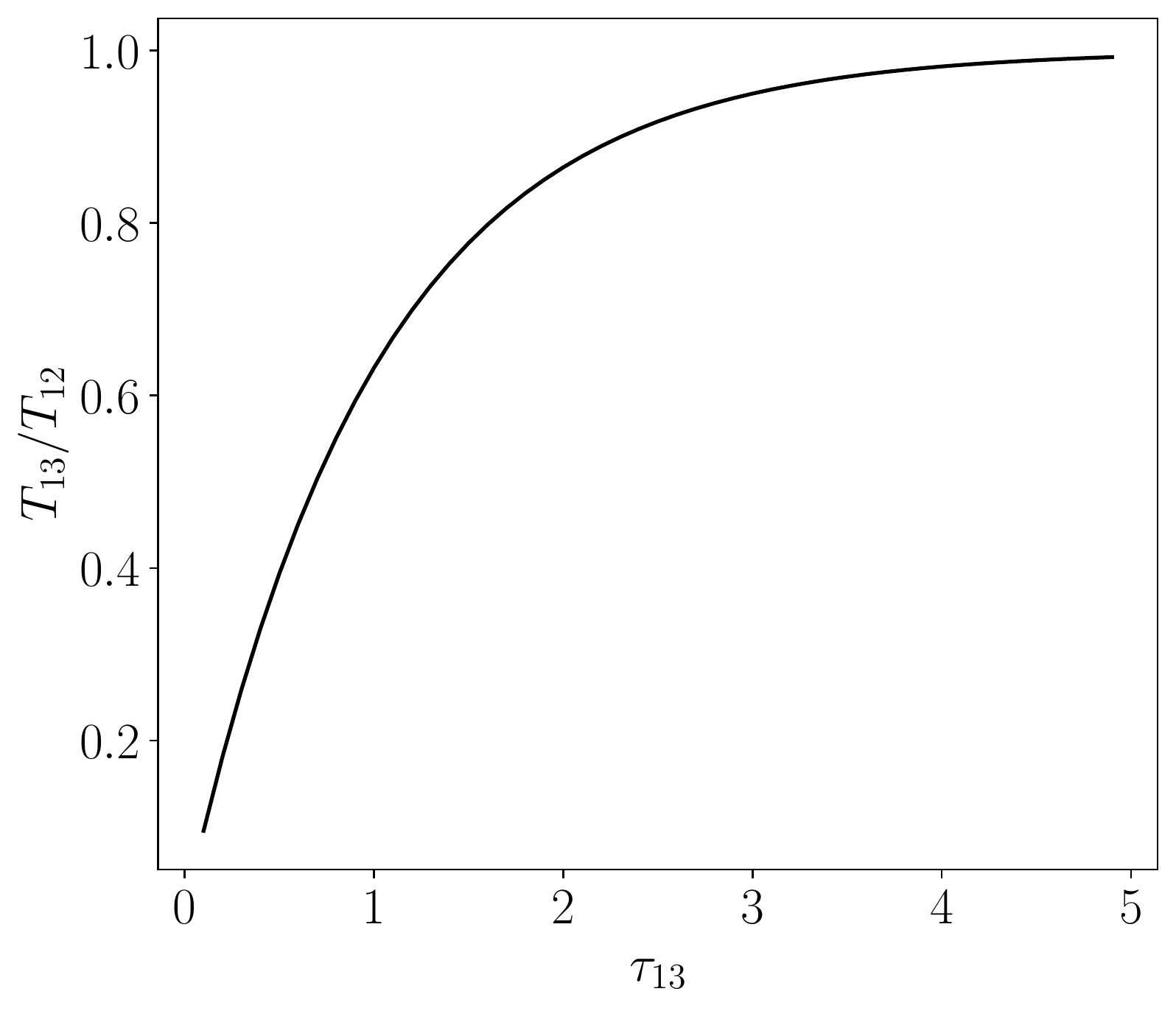}
\caption{\tco/CO line ratio as a function of \tco\ optical depth $\tau_{13}$ for the J=2-1 transitions. Here we assume the same excitation temperature for both lines and a \tco/CO abundance ratio of 1/69. For a line ratio of $\sim$0.2, the \tco\ optical depth is $\ll$1.}\label{fig:2}
\end{figure}

Figure \ref{fig:1} shows the spectra of CO and \tco~ beam-smoothed brightness temperature, averaged over each of the four mapped fields.  The ratio of \tco/CO peak intensities is in the range of 0.15 --- 0.25, which indicates that the \tco~ line is not optically thick, as illustrated in Figure \ref{fig:2}.  This conclusion will help justify our LTE analysis discussed below.  Three of the CO  spectra show single components of widths in the range 5 to 8 \kms, typical for GMCs.  The CO and \tco~ spectra for NGC~7538 are notably broader than the other clouds, though not with obvious multiple velocity components.  The line broadening in NGC~7538 is likely related to kinematic structure in the GMC, a result of the presence of several very massive O-stars possibly at different evolutionary stages with associated H~II regions \citep[cf.][and references therein]{2014MNRAS.439.3719C}.

\subsection{Peak and integrated intensity maps}

\begin{figure}
\plottwo{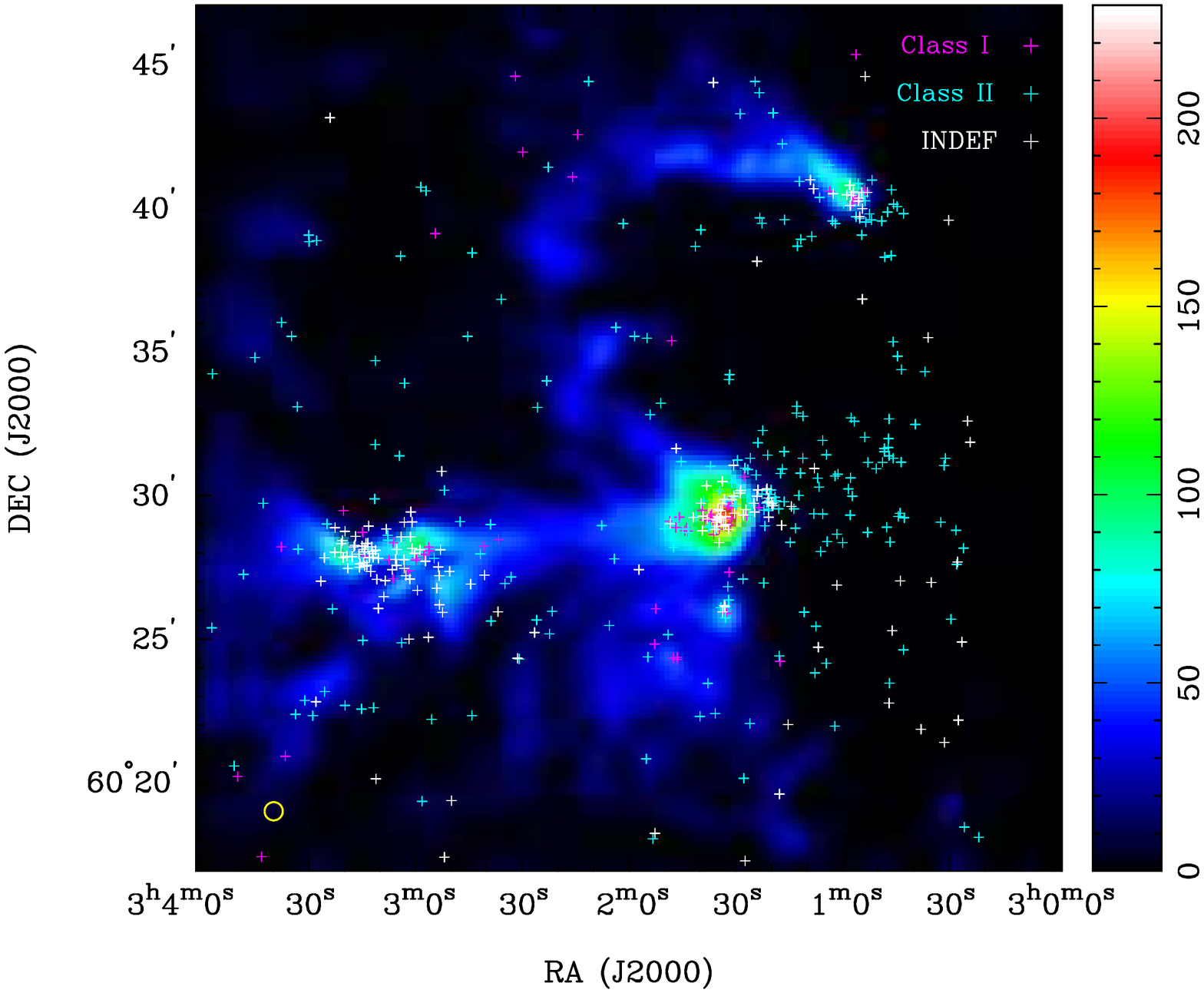}{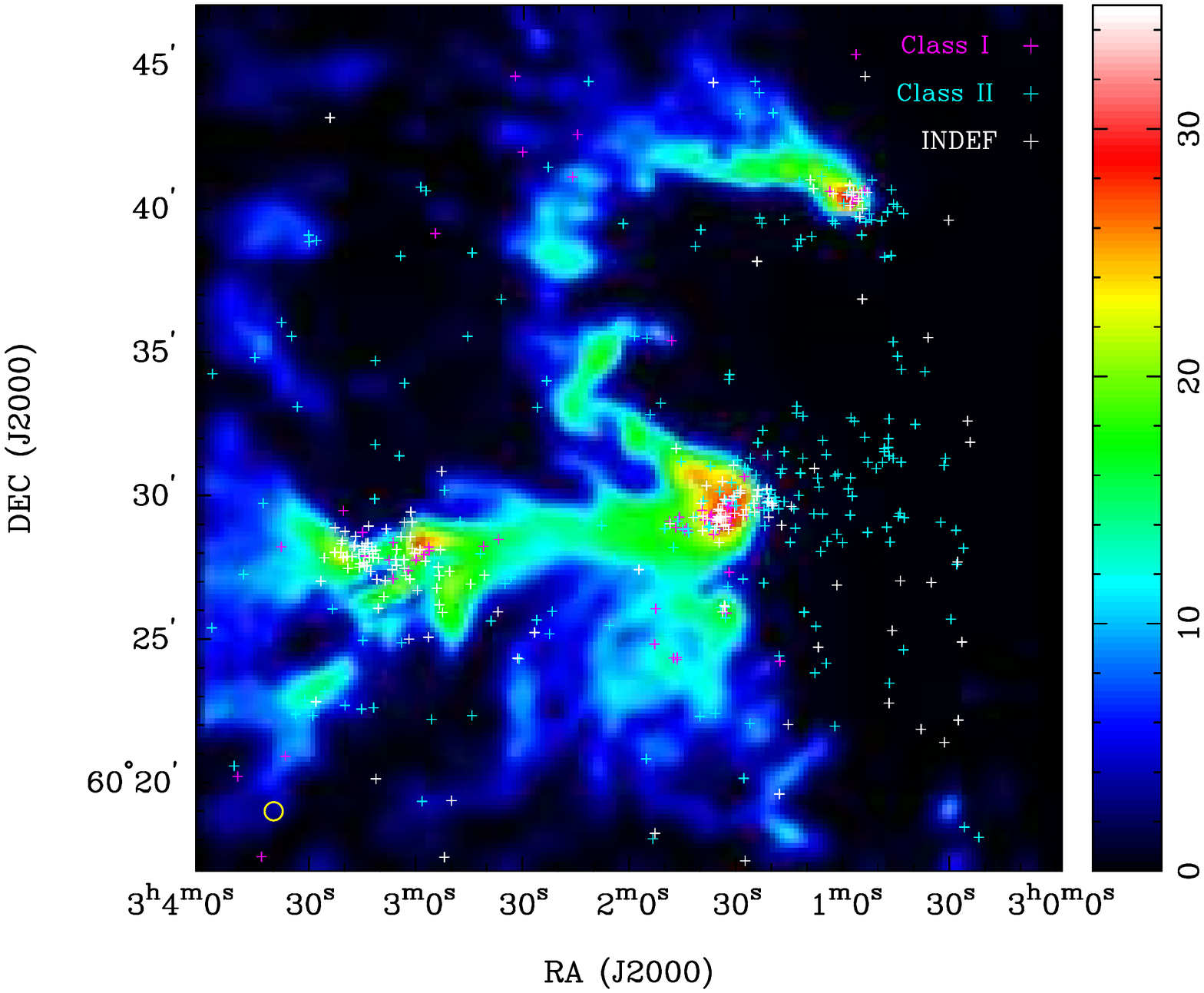}
\caption{CO J=2-1 integrated intensity distribution in K-\kms~(left panel) and peak brightness temperature in K (right panel) toward W5-east. Colored $+$~symbols mark the YSOs catalogued by \citet{2014MNRAS.439.3719C};  colors indicate their assigned YSO class, or ``Indefinite".  Yellow circle shows the map resolution of 38\arcsec (FWHM).}\label{fig:3}
\end{figure}

\begin{figure}
\plottwo{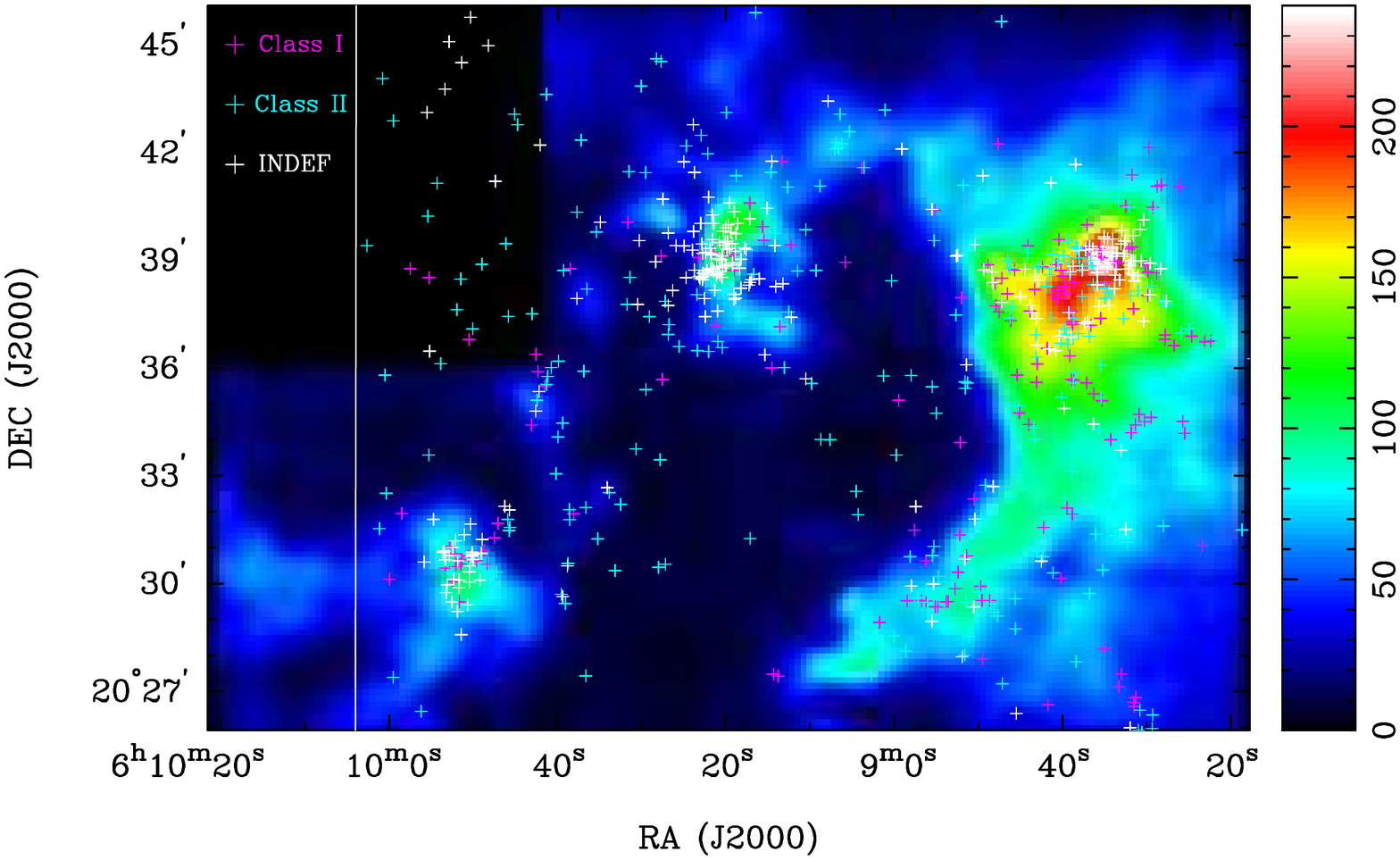}{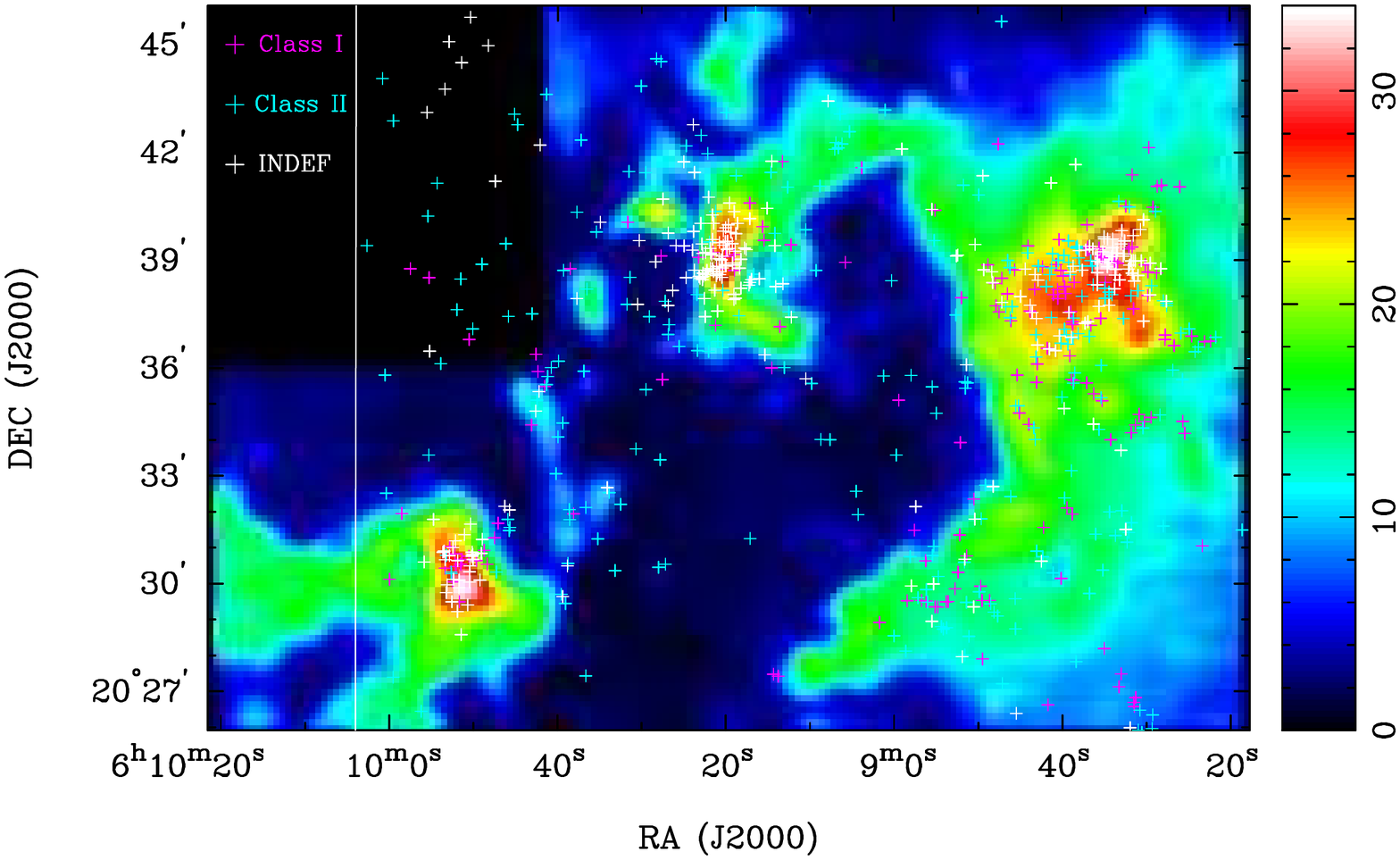}
\caption{CO J=2-1 integrated intensity distribution in K-\kms~(left panel) and peak brightness temperature in K (right panel) toward Sh2-252. Colored $+$~symbols mark the YSOs catalogued by \citet{2014MNRAS.439.3719C};  colors indicate their assigned YSO class, or ``Indefinite".  White line marks the eastern boundary of the region for which \citet{2014MNRAS.439.3719C} identified YSOs.}\label{fig:4}
\end{figure}

\begin{figure}
\plottwo{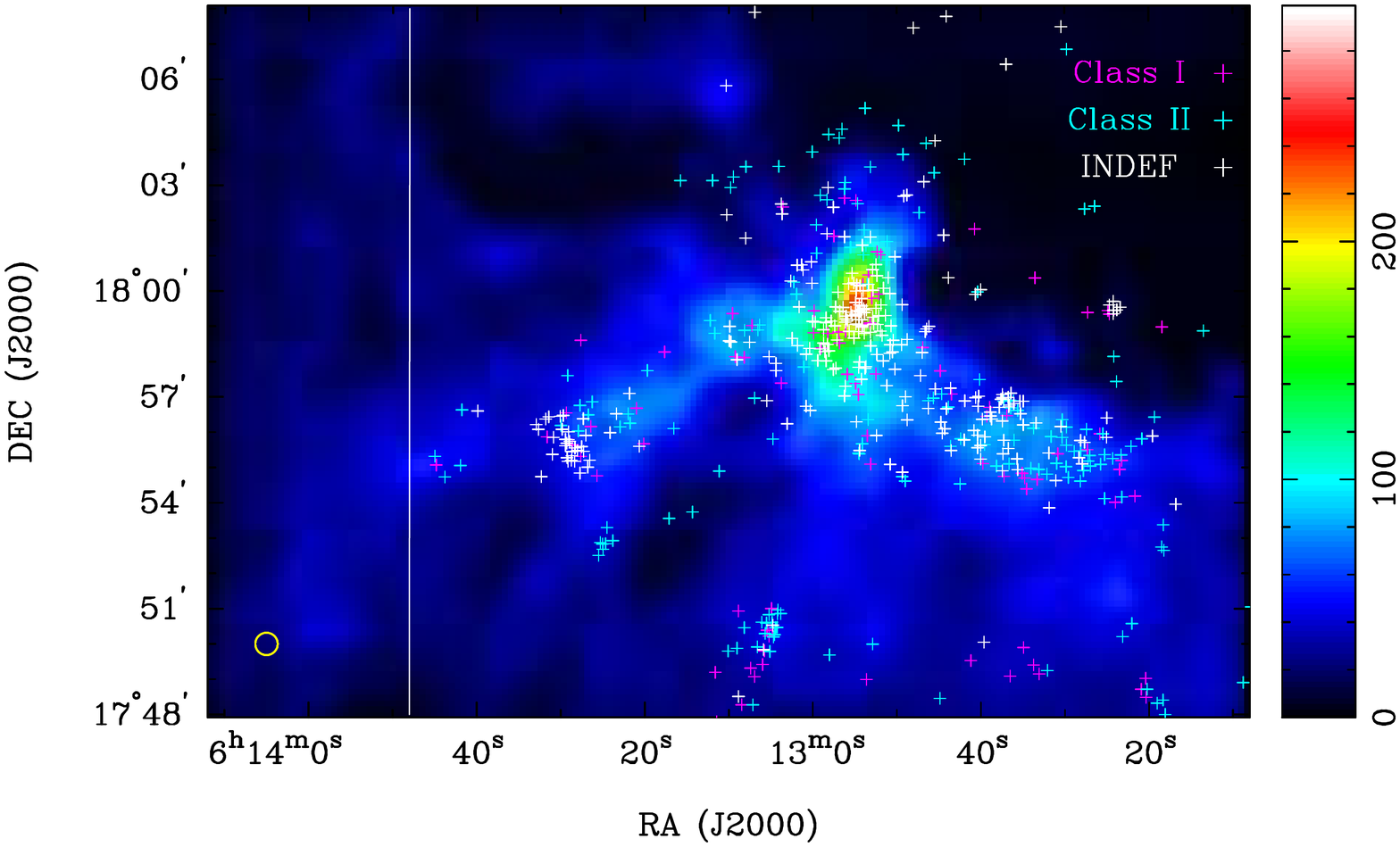}{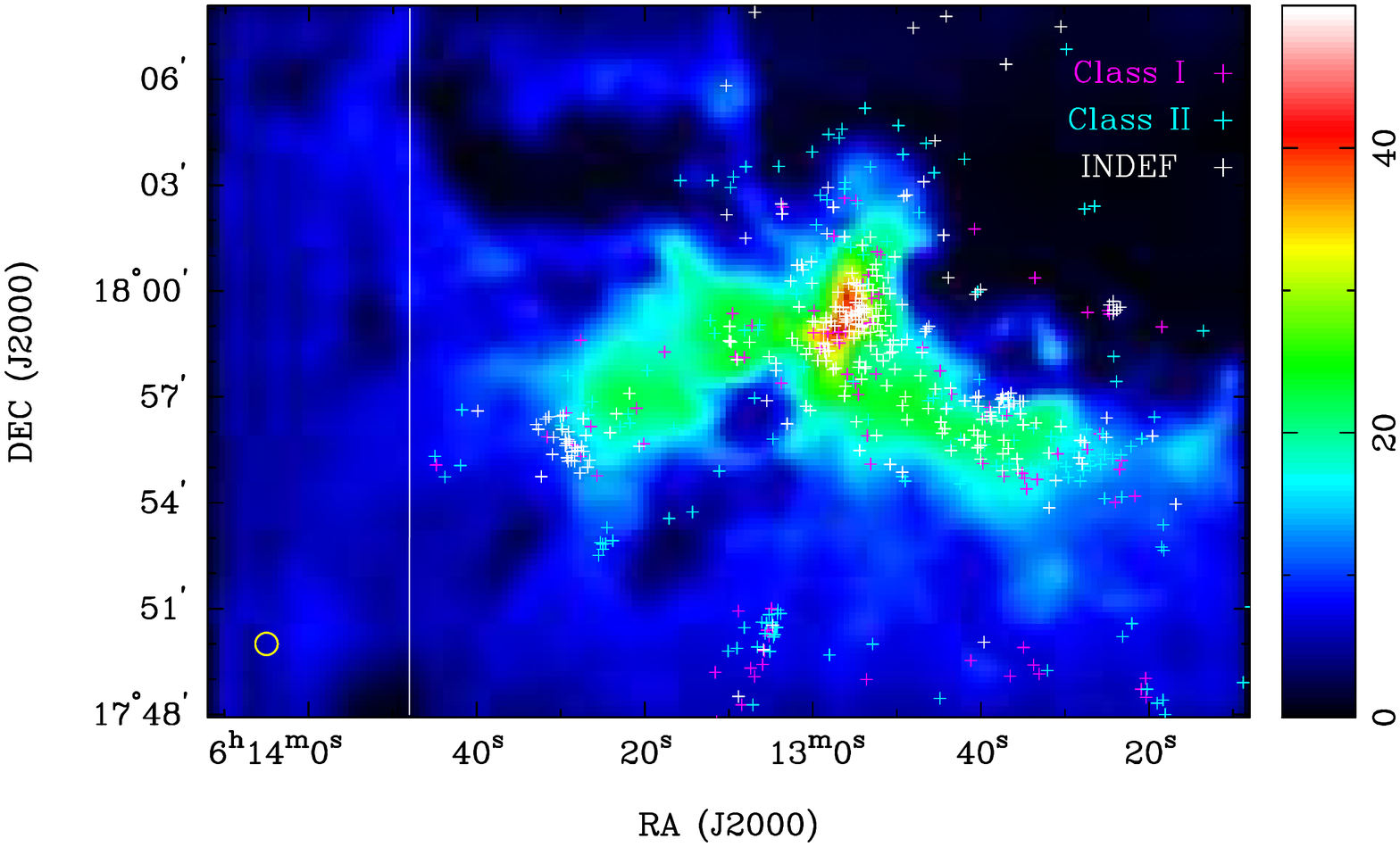}
\caption{CO J=2-1 integrated intensity distribution in K-\kms~(left panel) and peak brightness temperature in K (right panel) toward Sh2-254-258. Colored $+$~symbols mark the YSOs catalogued by \citet{2014MNRAS.439.3719C};  colors indicate their assigned YSO class, or ``Indefinite".  White line marks the eastern boundary of the region for which \citet{2014MNRAS.439.3719C} identified YSOs.}\label{fig:5}
\end{figure}

\begin{figure}
\plottwo{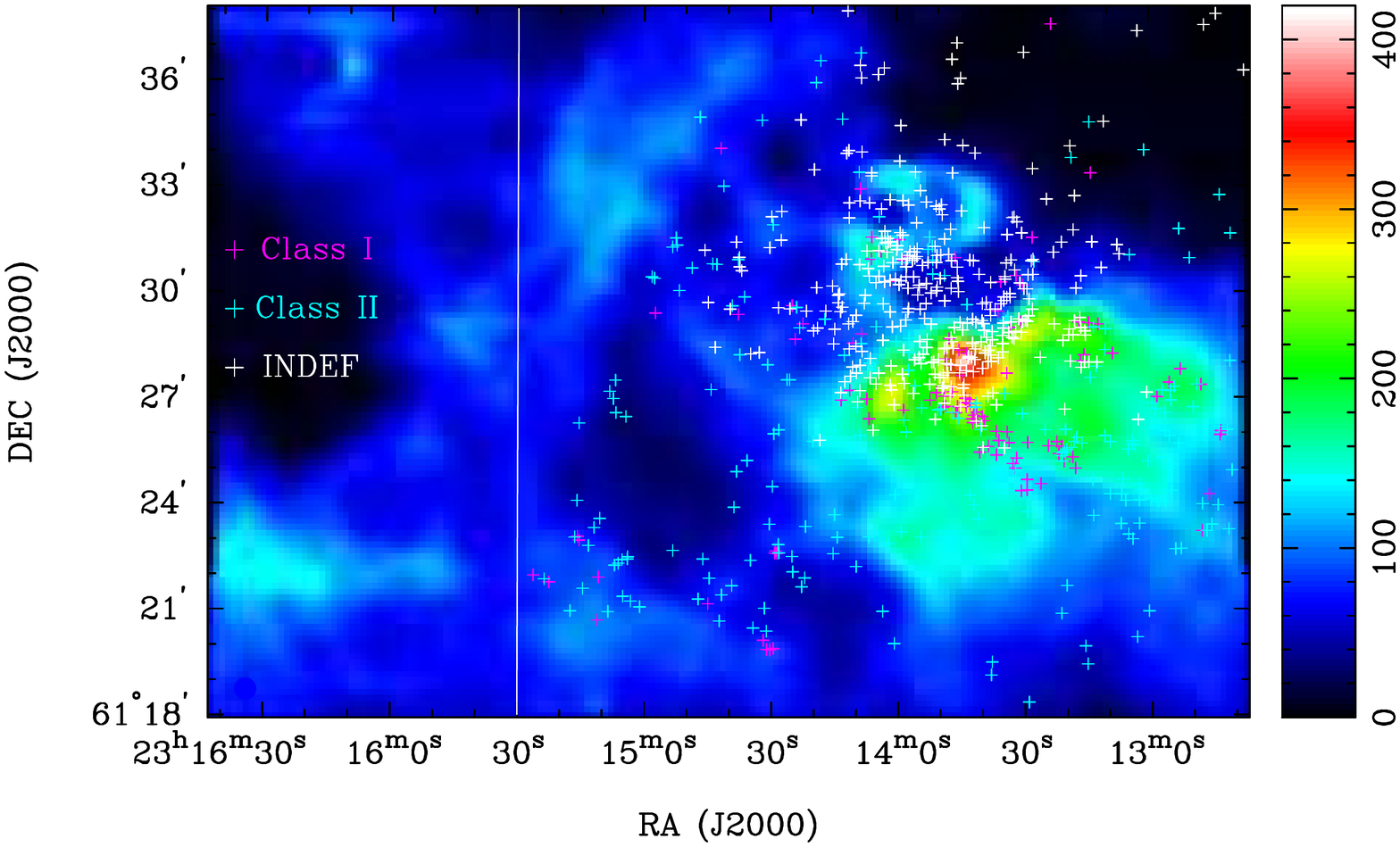}{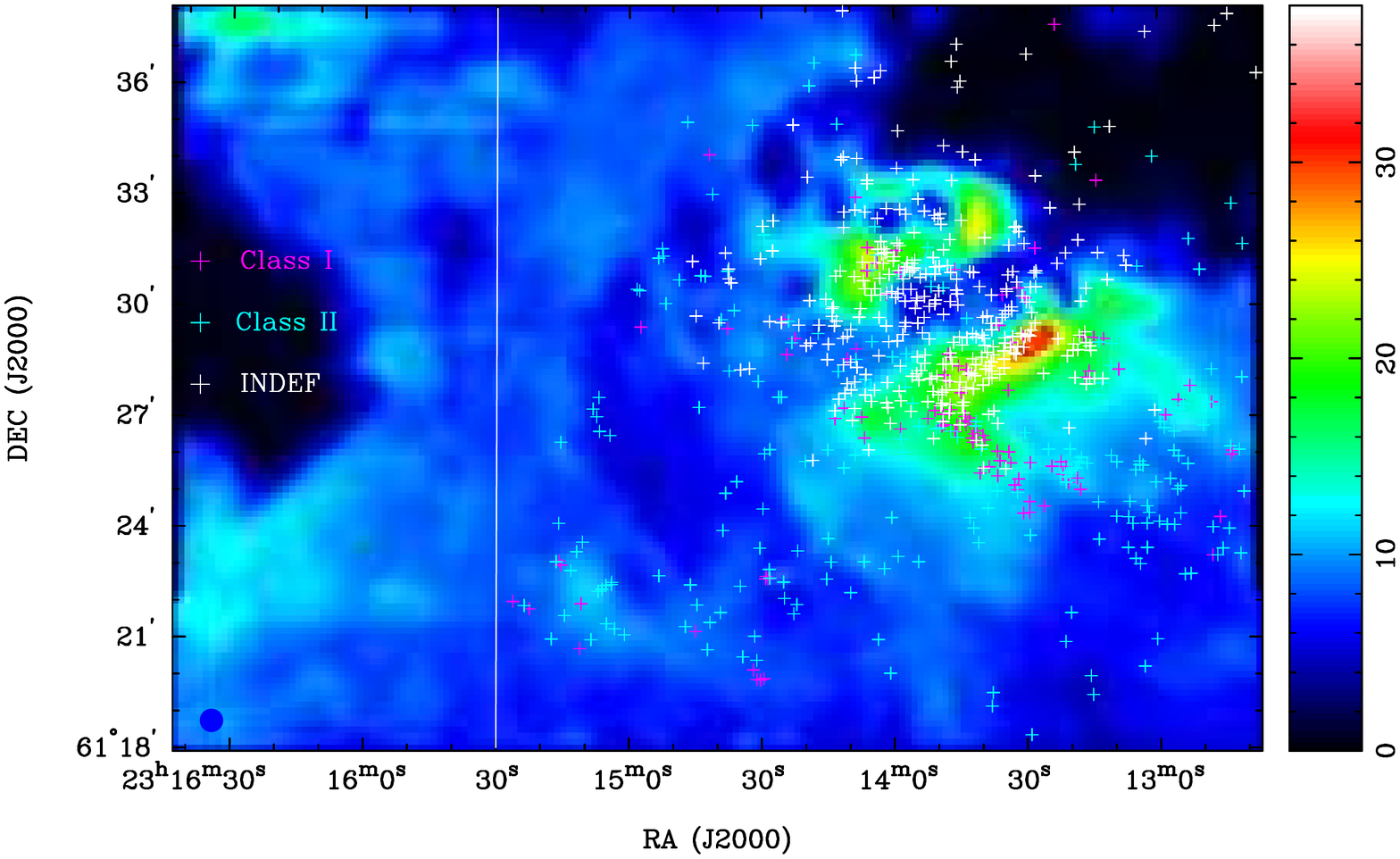}
\caption{CO J=2-1 integrated intensity distribution in K-\kms~(left panel) and peak brightness temperature in K (right panel) toward NGC~7538. Colored $+$~symbols mark the YSOs catalogued by \citet{2014MNRAS.439.3719C};  colors indicate their assigned YSO class, or ``Indefinite".  White line marks the eastern boundary of the region for which \citet{2014MNRAS.439.3719C} identified YSOs.}\label{fig:6}
\end{figure}

In Figures \ref{fig:3} through \ref{fig:6} we show maps of the integrated CO line brightness temperature in the left panels, and the maximum CO brightness temperature in the right panels.  The positions of YSOs catalogued by \citet{2014MNRAS.439.3719C} are indicated by colored crosses.  Class I stars are in magenta, Class II in cyan, and objects identified as YSOs but of ambiguous evolutionary class (``indefinite'') are white.  For all the GMCs except W5-east, our CO maps extended somewhat beyond the regions for which \citet{2014MNRAS.439.3719C} identified and catalogued YSOs.  The eastern boundary of their YSO maps is marked with thin white lines in the figures.  Generally the areas to the east of these lines show relatively little CO emission, and we do not include them in our comparison of LTE-derived gas column densities and YSOs presented below.

The CO J=2-1 line is optically thick over the bulk of all  GMCs, so the peak brightness temperature should be a measure of the excitation temperature of the transition.  Under typical cloud conditions, the CO rotational level populations up to J=3 are probably close to a thermal distribution (as discussed below).  Therefore the peak brightness temperature maps are expected to be good indicators of the gas kinetic temperature distribution in the clouds.

\subsection{LTE analysis: gas column density and CO excitation temperature distributions}

In our previous studies of Sh2-235 \citep{2016ApJS..226...13B} and Cep B/C \citep{2018ApJS..238...20B} GMCs, we also observed the J=3-2 transition of CO with resolution comparable to the J=2-1 maps over the same regions.  In both cases we found that the peak CO brightness temperatures in the two lines followed very closely the expected relation for an LTE population of the rotational levels, with the   J=3 level slightly subthermally excited, at $\sim$90\% of the kinetic temperature.  The J=1 and J=2 levels should therefore be populated very close to LTE conditions.  With that justification we applied an LTE analysis of the CO and \tco~ maps to derive the CO excitation temperature and column density distributions.  In the present work we do not have CO J=3-2 maps of the four GMCs, but these clouds appear to have properties very similar to those of Sh2-235, which was also included in the \citet{2014MNRAS.439.3719C} study of YSO content.  We therefore will assume that the CO rotational levels in these four GMCs are well described by LTE conditions as well.  

We have used the LTE analysis task {\it colden}  in the Miriad software package \citep{1995ASPC...77..433S} to derive maps of the CO excitation temperature and the \tco~column density, as described in \citet{2016ApJS..226...13B}.  Pixels where the CO emission is not detected (with a CO brightness temperature less than $5\times rms$)  are blanked in the maps.  The minimum CO intensity is cut off at 0.65 K brightness temperature per pixel in one 1 MHz spectral channel.

A critical physical effect that must be considered in deriving the total gas column density from the LTE-derived \tco~ column density is depletion of CO molecules onto dust grain surfaces in the coldest parts of the observed GMCs.  We have applied the recent results of \citet{Lewis_2021} to determine the assumed \tco~ abundance relative to molecular hydrogen, as a function of gas temperature.  \citet{Lewis_2021} used CO and \tco~ observations with an LTE analysis to determine the \tco~ gas phase column density distribution for sections of the California Molecular Cloud.  The CO observations and LTE analysis employed by \citet{Lewis_2021} were essentially identical to ours.  They also used {\it Herschel} far-IR maps with resolution comparable to our CO maps, to derive the mean grain temperature and the total column density of dust grains at each map pixel.  They applied a standard gas-to-dust ratio to determine the distribution of molecular hydrogen.  From these data they derived the \tco/H$_2$ gas phase abundance ratio, [\tco], as a function of dust temperature (their Figure 7).  We adopt a simple analytic fit to their results, namely

$$ [^{13}{\rm CO}]  =  (7.94\times10^{-11})  \times T^{3.10},  ~~~12 ~{\rm K} \leq T \leq 24~{\rm K} $$
$$ {[^{13}{\rm CO}]} =   1.51\times10^{-6}, ~~~24~{\rm K} < T$$

\noindent where $T$ is the gas temperature from the CO brightness temperature.  We assume the gas temperature equals the {\it Herschel}-derived dust temperature to convert the LTE \tco~ column density to the total molecular hydrogen column density at each map pixel.  We further assume that H$_2$~ is the only form of hydrogen in the mapped regions.  For $T < 12$~K, the depletion correction is uncertain, so these pixels are blanked in the column density maps.  To account for the presence of He and CNO elements we multiply the H$_2$ column density by a factor of 1.40 \citep[cf.][]{2011piim.book.....D}, yielding the total gas column density, $\Sigma_{gas}$.   After applying the depletion correction and a factor 1.4 for heavier elements, we find that the maps are  sensitive down to a minimum $\Sigma_{gas} \ge 40 $ \msun~pc$^{-2}$ (corresponding to $A_V \approx 2.5$~mag for standard gas/dust values) for all four GMCs.

Results are presented in Figures \ref{fig:7} - \ref{fig:10}, which show the total gas column density (expressed in units of \msun~pc$^{-2}$) in the left panels, and the CO excitation temperature in Kelvins in the right panels.  

In Table \ref{tab:derive} we summarize the total molecular gas mass for each GMC, as derived from our LTE analysis including the correction for depletion of CO.  Only unblanked pixels are included in the total mass.  The mean gas column density in the last column of Table \ref{tab:derive} is total gas mass divided by the area of the unblanked pixels, not the full area mapped.  The mean column densities for gas with detected CO emission range from 167 to 487 \msun~pc$^{-2}$.  These values are substantially higher than the mean GMC surface density of 35 \msun~pc$^{-2}$ derived by  \citet{2020ApJ...898....3L} for galactic GMCs.  Their values, however, are calculated for the full extent of the GMC out to the minimum detectable level of CO J=1-0 emission, while our maps are limited to the cloud region where YSOs are cataloged, i.e., where star formation is currently most active, which does not cover the full extent of detectable CO emission.  Moreover, their cloud mean surface densities are based on analyses using adopted values of X$_{\rm CO}$, the nominal conversion factor from CO line intensity to total gas column density, which do not (explicitly) consider the effects of temperature-dependent CO depletion onto grains.  It is therefore not surprising that the average column densities in Table \ref{tab:derive} are considerably higher than 35 \msun~pc$^{-2}$.


In general, all of the maps of molecular gas column density have the greatest part of the cloud area with relatively low $\Sigma_{gas}$, below 200 \msun~pc$^{-2}$.  The highest column density areas cover only a small fraction of the mapped field, and are invariably associated with a high surface density of YSOs.  In contrast, the maps of CO excitation temperature (which should be close to the gas kinetic temperature for the conditions of gas density expected to exist in these clouds) have a more uniform distribution, with values typically in the range 10 - 20 K.  Pixels with higher temperatures, $\geq 30$~K, are confined to relatively smaller areas.  Note that the areas of highest gas column density are not necessarily the same as those with the highest temperatures.  The latter regions often appear to lie at the edges of associated H~II regions.  

\subsection{Comparison with associated YSOs and H~II regions}\label{subsec:compare}

The GMCs in this study have been known previously as optical H~II regions or radio continuum sources of free-free emission from ionized gas.  The ionizing photons are radiated by newly formed OB stars, whose energy input is affecting the molecular gas by physical processes referred to as ``feedback" dominated by photoionization and stellar winds for GMCs which have not yet produced a supernova (SN).  None of the clouds in this study shows evidence for a SN event within the cloud lifetime, neither by an optical remnant, nor X-ray or non-thermal radio continuum emission.  The lack of evidence for recent supernovae is consistent with the relative youth of the embedded YSO clusters, which have not yet had time to evolve to core collapse, i.e., the ages of the clusters are $\leq 3$~Myr.  

\begin{figure}
\plottwo{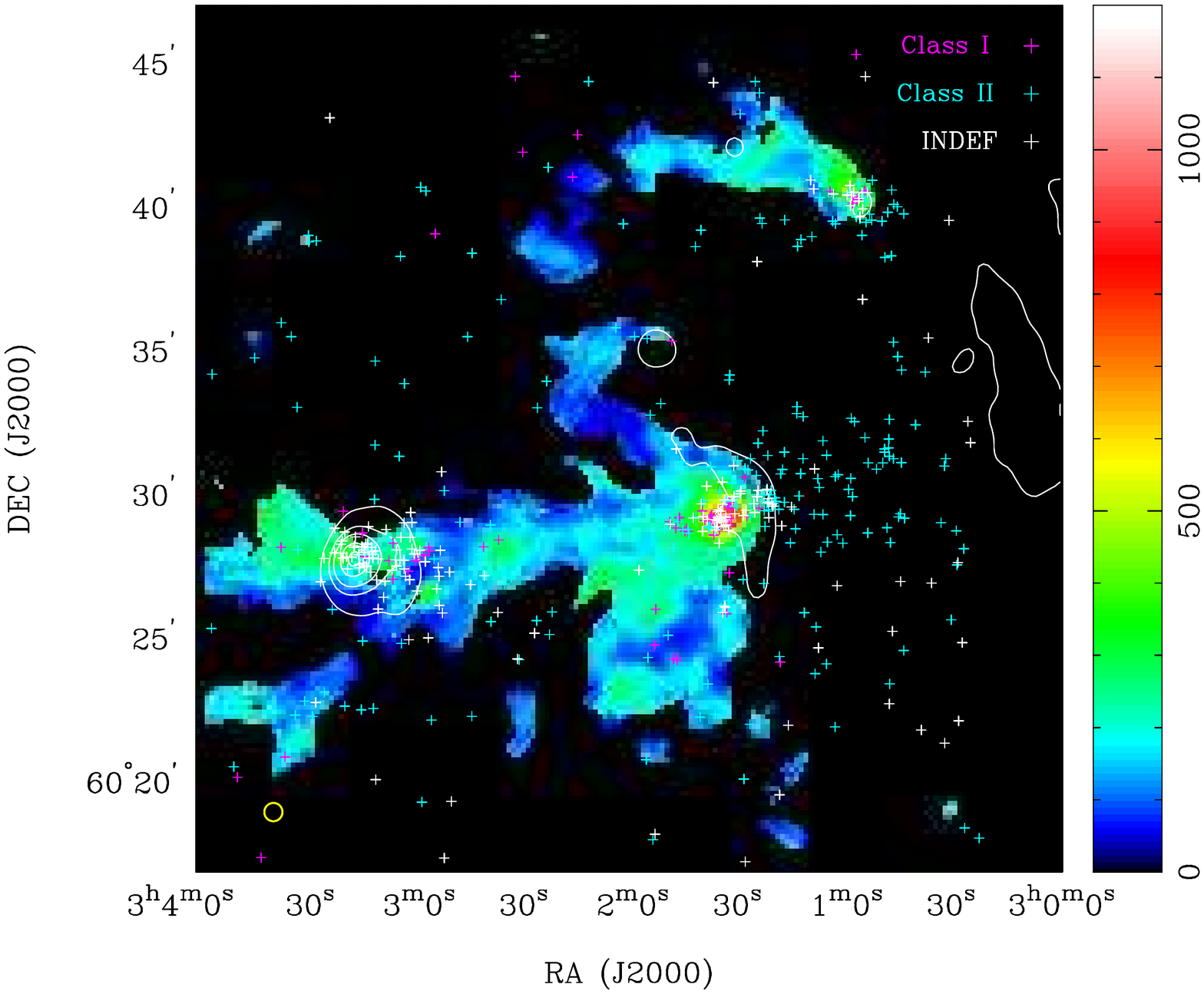}{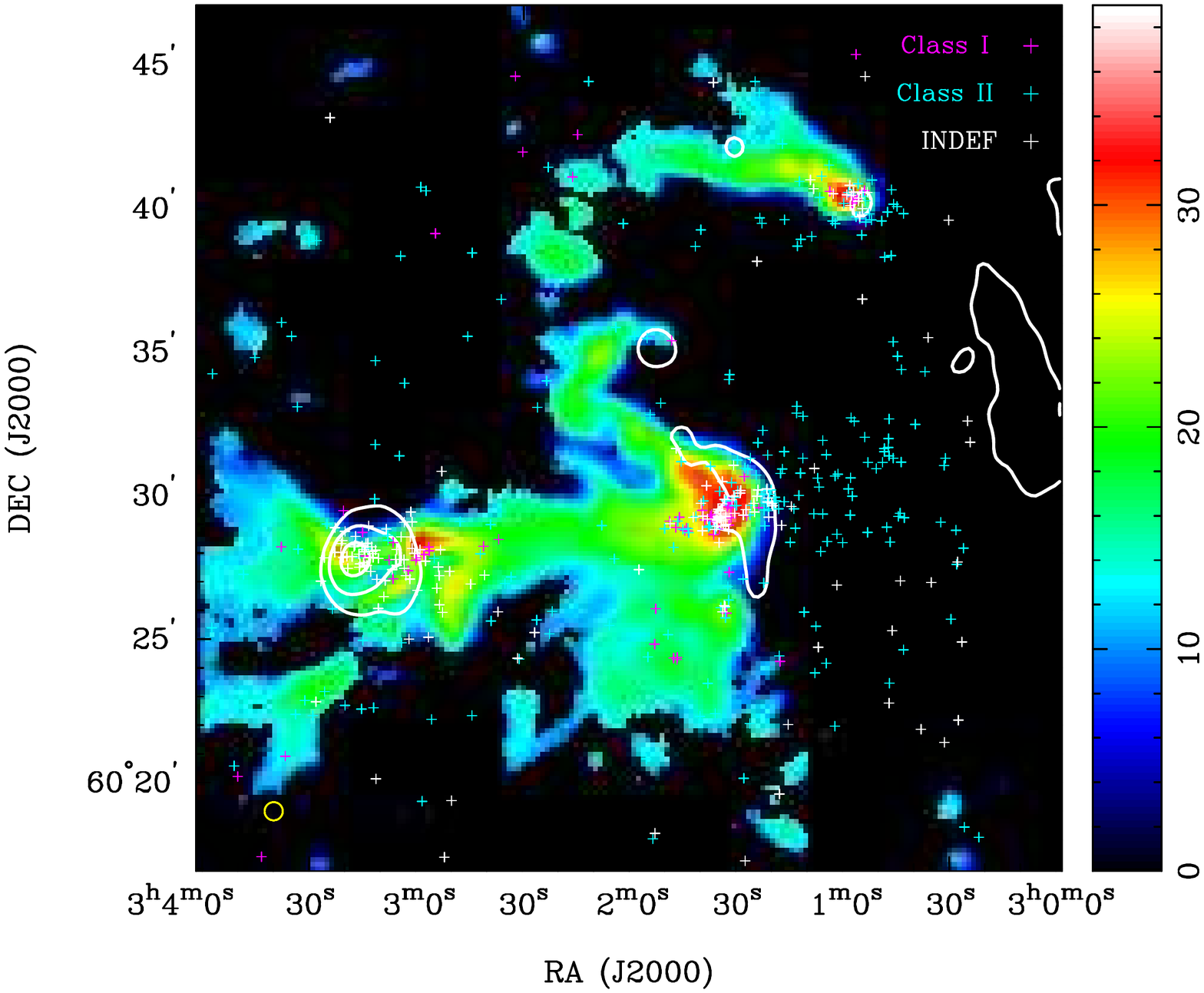}

\caption{Total molecular gas column density in \msun ~pc$^{-2}$ (left panel) and CO J=2-1 excitation temperature in K (right panel) for W5-east, derived from LTE analysis with correction for CO depletion (see text).  White contours show 1.4 GHz continuum brightness temperature from the Canadian Galactic Plane Survey \citep[CGPS,][]{Taylor_2003}, with levels at 10, 20, 30, 50, and 70 K.  Colored crosses mark positions of YSOs from \citet{2014MNRAS.439.3719C} as in Figure \ref{fig:3}.}
\label{fig:7}
\end{figure}

\begin{figure}
\plottwo{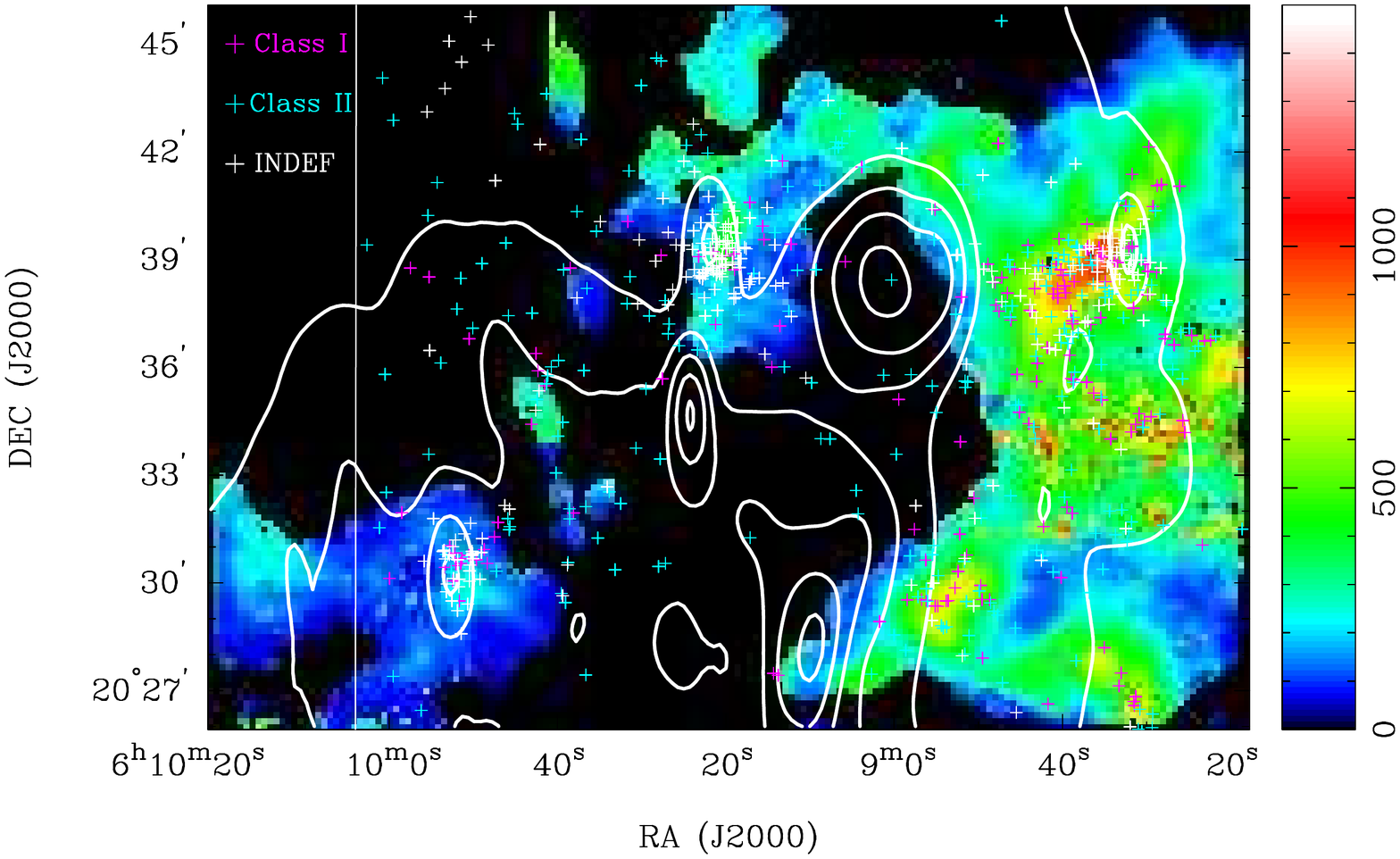}{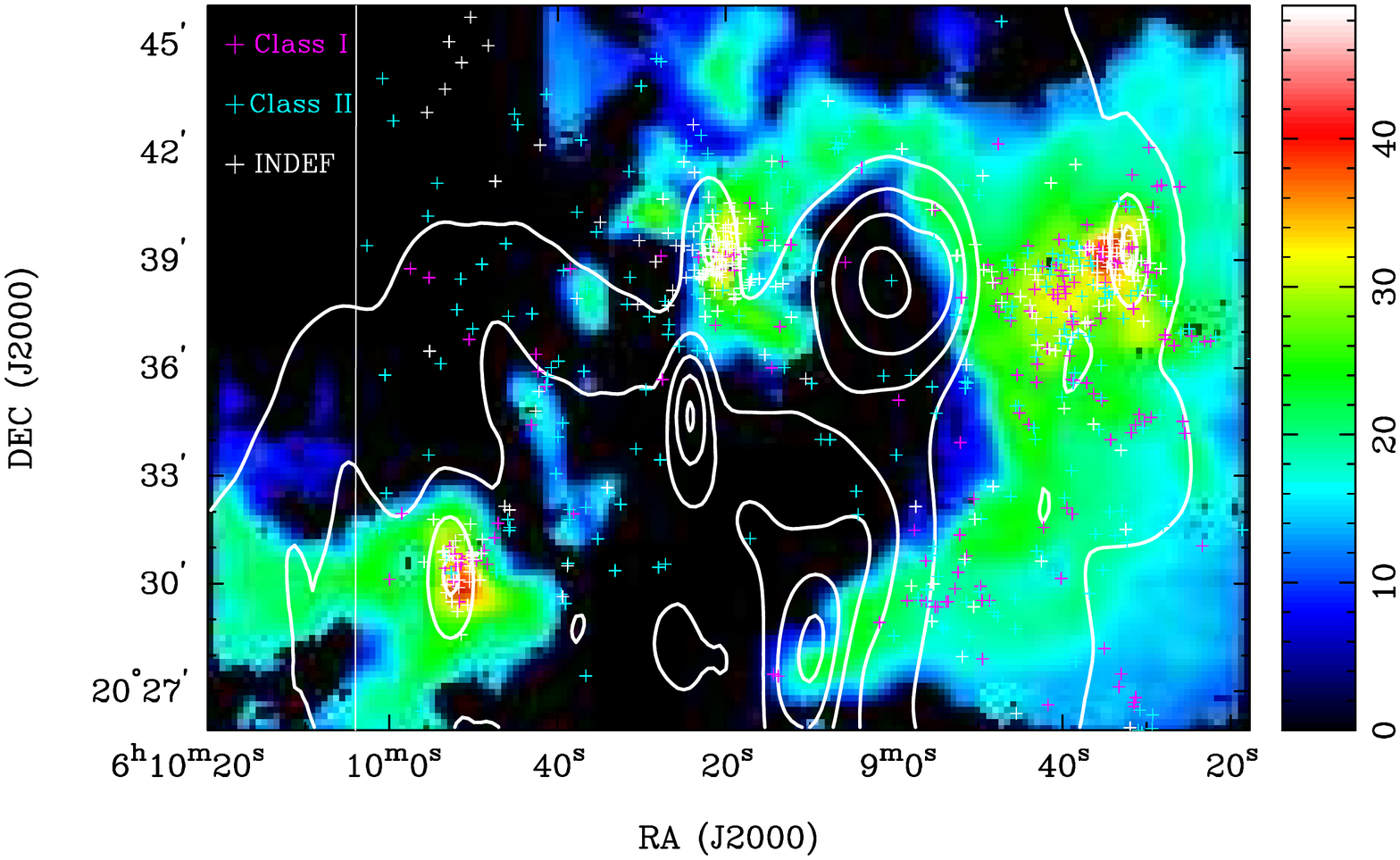}
\caption{Total molecular gas column density in \msun ~pc$^{-2}$ (left panel) and CO J=2-1 excitation temperature in K (right panel) for Sh2-252, derived from LTE analysis with correction for CO depletion (see text).  White contours show 1.4 GHz continuum brightness temperature from the CGPS \citep{Taylor_2003}, with levels at  5, 10,  15, 20, 25, and 30K.  Colored crosses mark positions of YSOs from \citet{2014MNRAS.439.3719C} as in Figure \ref{fig:4}.}\label{fig:8}
\end{figure}

\begin{figure}
\plottwo{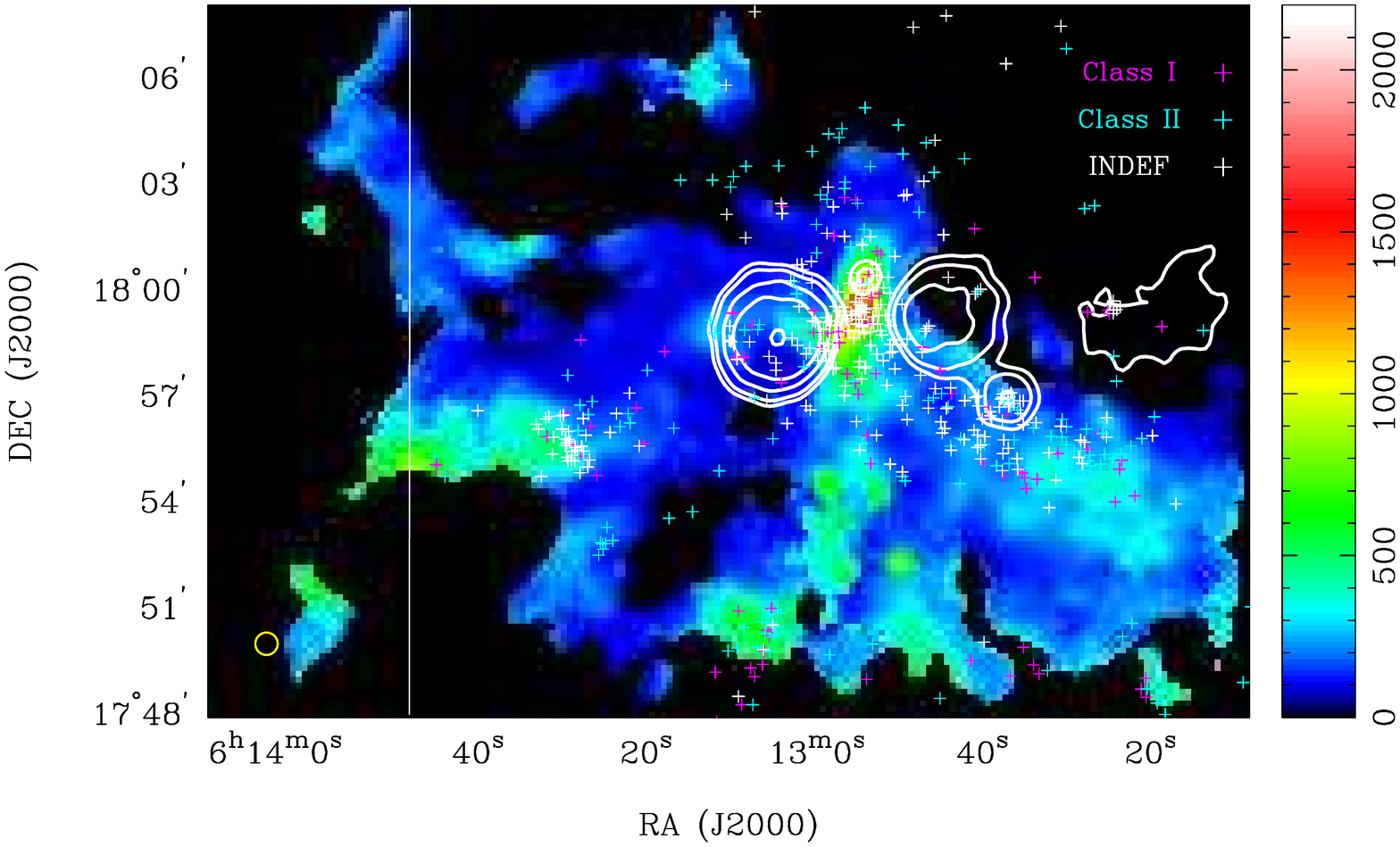}{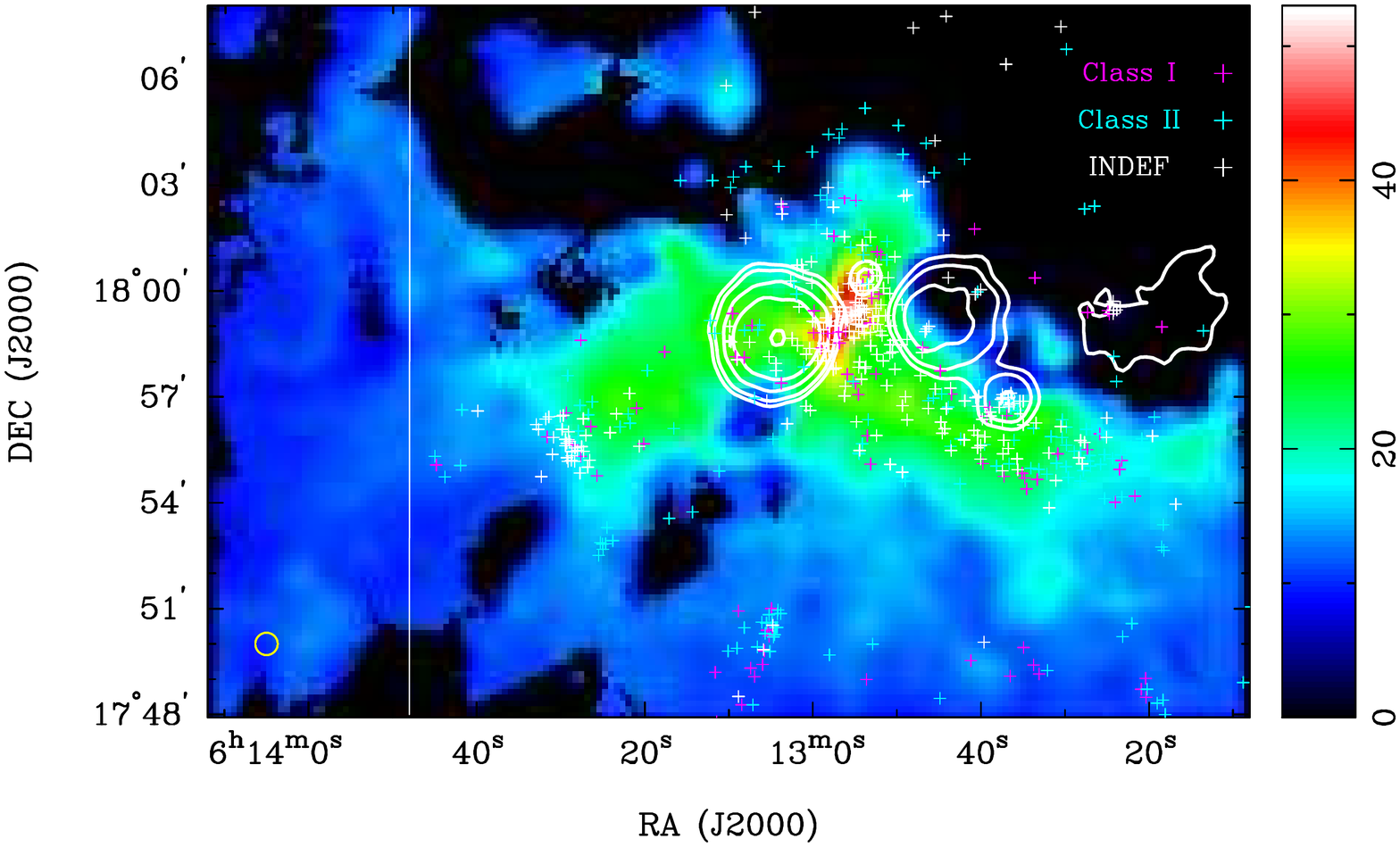}
\caption{Total molecular gas column density in \msun ~pc$^{-2}$ (left panel) and CO J=2-1 excitation temperature in K (right panel) for Sh2-254---258, derived from LTE analysis with correction for CO depletion (see text).   White contours show 1.4 GHz continuum surface brightness from VLA observations by \citet{1993ApJS...86..475F}, with levels at 0.01, 0.02, 0.05, 0.1, and 0.2 Jy/beam.  Colored crosses mark positions of YSOs from \citet{2014MNRAS.439.3719C} as in Figure \ref{fig:5}.}\label{fig:9}
\end{figure}

\begin{figure}
\plottwo{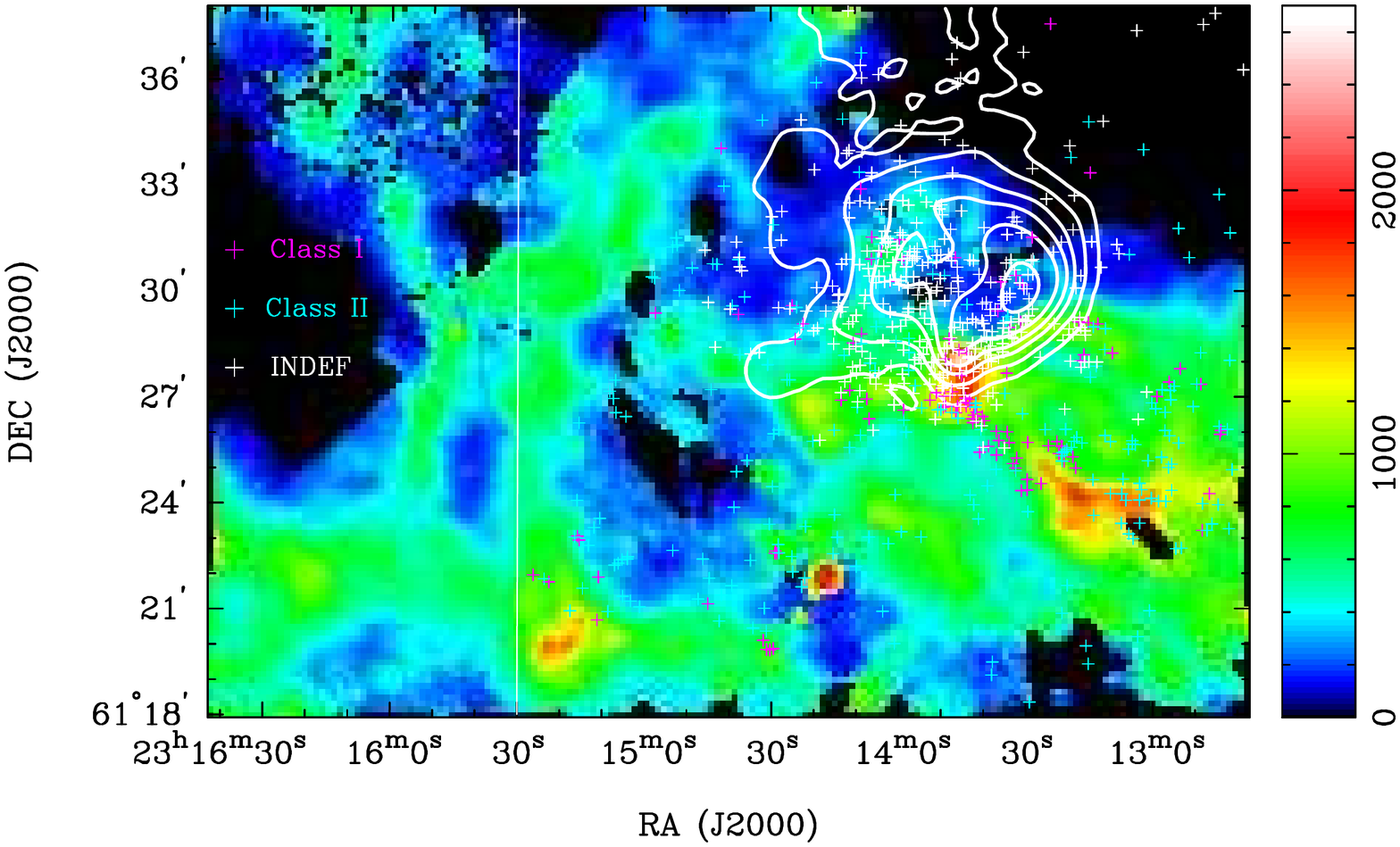}{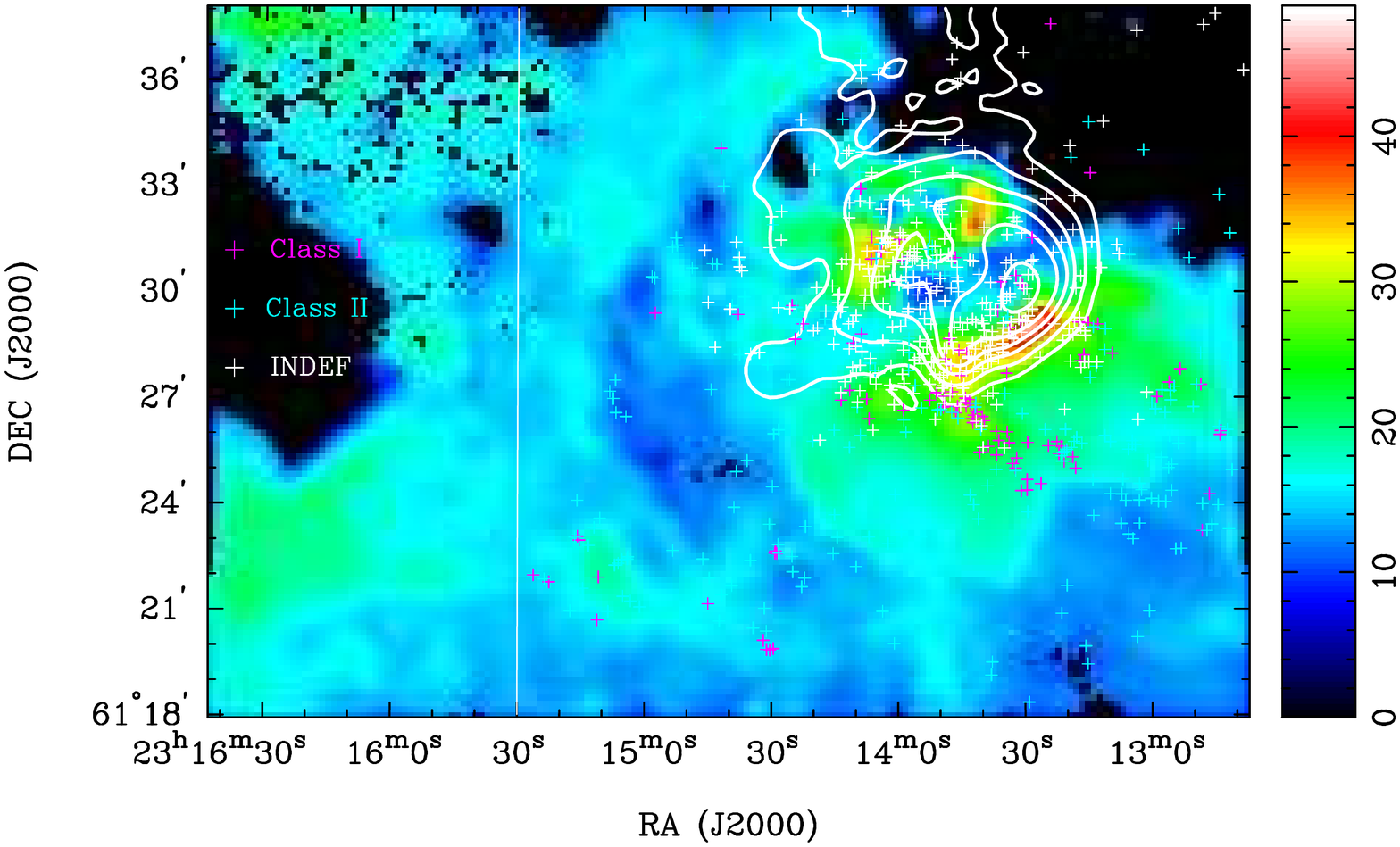}

\caption{Total molecular gas column density in \msun ~pc$^{-2}$ (left panel) and CO J=2-1 excitation temperature in K (right panel) for NGC~7538, derived from LTE analysis with correction for CO depletion (see text).  White contours show 1.4 GHz continuum brightness temperature from the CGPS \citep{Taylor_2003}, with levels at  10, 20, 50, 100, 200, and 500K.  Colored crosses mark positions of YSOs from \citet{2014MNRAS.439.3719C} as in Figure \ref{fig:6}.}\label{fig:10}
\end{figure}

In Figures \ref{fig:7} - \ref{fig:10} we also show as white contours the radio continuum emission at $\lambda$21 cm that was detected in the Canadian Galactic Plane Survey \citep[CGPS,][]{Taylor_2003} for W5-east, Sh2-252, and NGC~7538, or by the VLA \citep{1993ApJS...86..475F} for Sh2-254--258.  These radio sources all appear to arise from free-free emission radiated from gas ionized by the newly formed OB stars.  At this wavelength dust extinction should be negligible so the radio continuum reveals the location of photoionization within the molecular clouds.  

The CGPS images may not show regions of diffuse low surface brightness continuum such as would come from older, extended H~II regions.  In W5-east  (Figure \ref{fig:7}) there is a relatively bright, compact source at a minor peak in the molecular column density, and an elongated weak source that appears to wrap around the west side of the highest molecular peak.  There is a high concentration of YSOs to the west but with no detected CO within several arcminutes of the stellar cluster.  The absence of molecular gas suggests that this region has been cleared out by a combination of stellar winds and photoionization from stars in the YSO cluster.  The appearance of the molecular gas also looks as though the neutral gas has been compressed and swept into elongated structures by stellar feedback mechanisms.  

The radio continuum toward Sh2-252 (Figure \ref{fig:8}) has three strong resolved peaks that lie in a molecular void which may have resulted from photoionization or expansion of the H~II region.  The molecular gas to the west of the main radio continuum appears to have been shaped and compressed by such expansion.  In contrast to W5-east, there is no YSO cluster in the molecular void, so the ionization source is not evident in the distribution of young stars.

The group of H~II regions Sh2-254--258 (Figure \ref{fig:9}) was discussed extensively by \citet{2009AJ....138..975B} and by \citet{2008ApJ...682..445C}.  Our new CO and \tco~ maps confirm these earlier results.  The VLA map of radio continuum emission reveals two bright, nearly spherical H~II regions (Sh2-255 and Sh2-257) on either side of the highest molecular column density peak, which appears to have been compressed and heated by the expansion of the ionized gas on each side.  To the northwest, a void in the CO coincides with the largest of the H~II regions, Sh2-254, which has swept away or ionized the molecular gas.  \citet{2009AJ....138..975B} proposed that this was an example of sequential formation of the stars exciting the H~II regions,
triggered by the compression of the molecular gas as a consequence of the expansion of the adjacent H~II regions. 

The radio continuum toward NGC~7538 (Figure \ref{fig:10}) has the morphology expected of an H~II region expanding to the northeast and compressing an adjacent molecular cloud to the southwest.  The highest molecular column density ridge lies at the southern edge of the ionized gas, with the molecular peak lying at a local extension of the radio continuum, seen as the white contours protruding to the south.  This extension may indicate the presence of a recently formed compact H~II region within the molecular peak and energized by a young OB star.  The corresponding map of CO excitation temperature---which we equate to the gas temperature---seen in the right panel of Figure \ref{fig:10}, shows that the molecular gas lying at the perimeter of the ionized gas is being heated to the range 30 to 40 K, presumably by radiation from the H~II region.  

The appearance of the molecular gas juxtaposed with the free-free continuum and the embedded YSO clusters (Figures \ref{fig:7} - \ref{fig:10}) can be compared to the results of numerical simulations which include feedback effects from photoionization and stellar winds.  An example is the work of \citet{2014MNRAS.442..694D} who present high spatial resolution 3D SPH numerical simulations of star-forming molecular clouds with both photoionization and stellar winds.  Several of their models have masses comparable to the GMCs in this study, and cover timescales that are comparable to the expected ages of the star formation activity detected in these clouds.  The morphology of the molecular gas and the location of the H~II regions we observe bear a strong resemblance to their simulations.   Typically a cavity of ionized gas forms within the molecular cloud.  \citet{2014MNRAS.442..694D} find that in most cases the dynamical effects of the expanding hot photoionized gas dominate over the effects of stellar winds, except at very early times in their simulations.  Their Figure 8 showing run I at 2.2 Myr including both stellar winds and photoionization, resembles our image of NGC~7538 (Figure \ref{fig:10}) fairly well.  The central cavity in the molecular gas is occupied by the H~II region and one side is compressed to high column density where clusters of YSOs are detected.  \citet{2014MNRAS.442..694D} do not explicitly include star formation in their simulations but the location of pixels with very high column density indicates where stars are expected to form.  A group of high density pixels centered in the ionized cavity implies a stellar cluster that provides the ionizing photons.  Two other groups of high density pixels lie on the rim of molecular gas surrounding the cavity, tracing the locations where more YSO clusters are likely to form.  We see similar morphologies in the molecular and ionized gas distributions in NGC~7538.  Broadly speaking, then, the predictions of these numerical simulations are consistent with the appearance of the molecular clouds in our study, and with their relationship to the ionized gas and locations of embedded YSO clusters as illustrated in Figures \ref{fig:7} - \ref{fig:10}.

\section{Discussion}

\subsection{Star formation rate vs. molecular gas column density}\label{subsec:sfrsigma}

The maps of total molecular gas column density, with the locations of the identified YSOs indicated by $+$ symbols (Figures \ref{fig:7} - \ref{fig:10}, left panels) show that the regions of high gas column density invariably have a strong concentration of YSOs associated with the gas.  (The converse is not always the case, however, as discussed in section \ref{subsec:compare}.)  To quantify this relation, we use the following method. 

\begin{figure}
\plotone{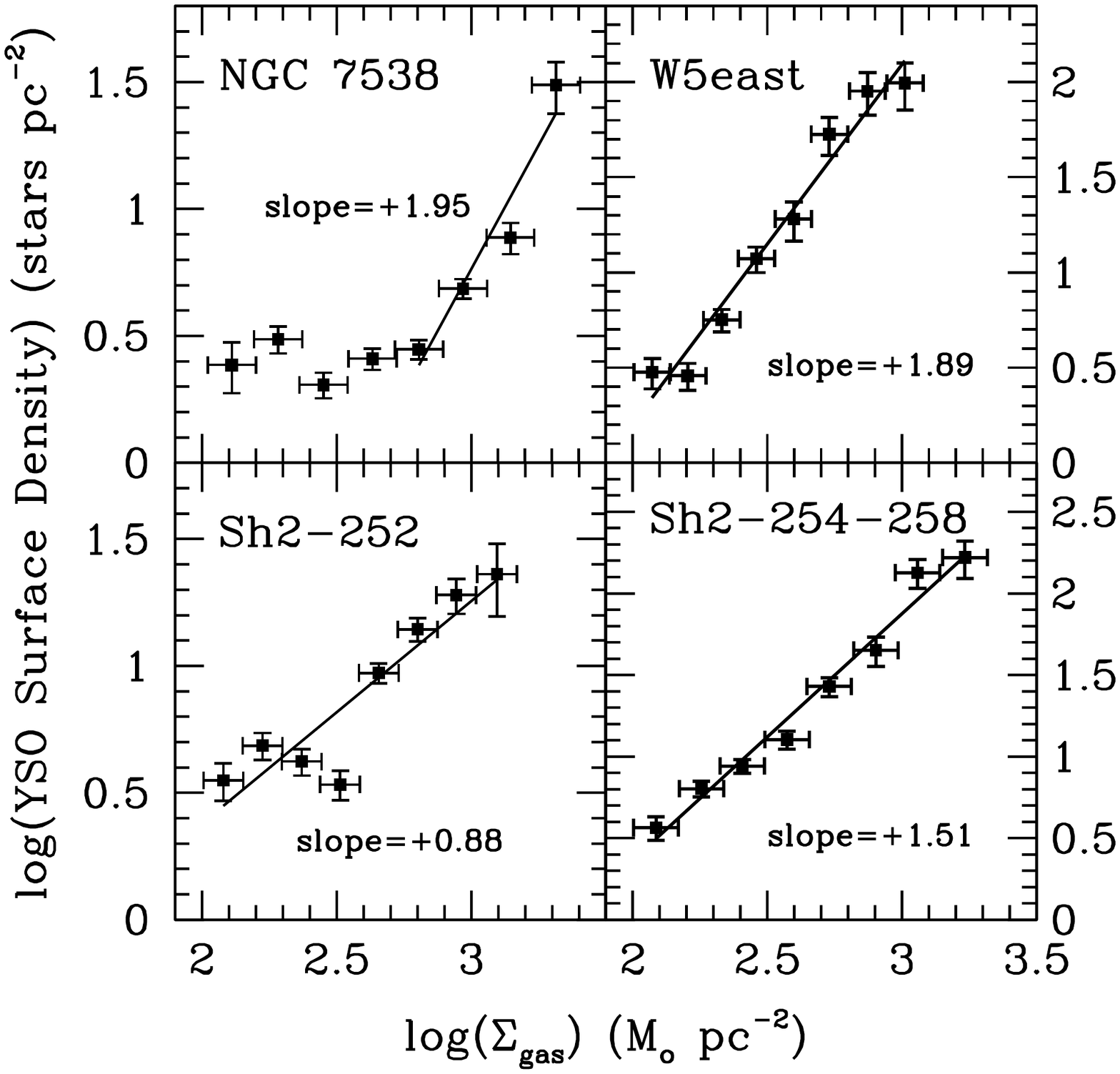}
     \caption{Mean YSO surface density for pixels with total molecular gas column density ($\Sigma_{gas}$) within bins of width shown by the horizontal bars.  Vertical error bars show standard deviations based on Poisson statistics for the number of YSOs counted within each $\Sigma_{gas}$~ bin.  Straight lines show fits to the plotted points (log-log space).  Slope gives the power-law exponent of the fit.  For NGC~7538, the fit is only to the highest 4 points.  Note that the scales differ between panels for the vertical axes.}\label{fig:11}
\end{figure}

First we create a set of pixel masks which select only pixels in the gas column density maps having values of $\Sigma_{gas}$ within a specified range.  The masks are defined for 8 bins which are spaced equally in log($\Sigma_{gas}$) from 100 \msun~pc$^{-2}$ up to the maximum map value.  This choice of widths ensures that each bin contains  enough YSOs for a statistically significant sample.  Then we apply each mask successively to the corresponding catalog of YSOs, counting the number of stars within the unmasked pixels.  We divide the number of YSOs by the area enclosed by the unmasked pixels to find the mean value of YSOs per pc$^2$ for each bin of $\Sigma_{gas}$.  The results are shown in Figure \ref{fig:11} as a log-log plot.  The horizontal bars show the width of each bin; the vertical error bars show the Poisson 1-$\sigma$ uncertainty based on the number of YSOs within the range of the bin.  For three of the GMCs, the points are reasonably well-described by a linear relation, implying a power-law dependence of the mean YSO surface density on the gas column density.  For NGC~7538, the highest 4 bins are close to a linear relation, but the 4 lowest bins show nearly a constant value of YSO surface density.   We perform a linear regression for each panel and show the slope of the fit (the power-law exponent) as well in the Figure.  For NGC~7538, the fit is only to the upper 4 bins.  

The observed current surface density of YSOs can be used to estimate the rate of star formation (in \msun~My$^{-1}$~pc$^{-2}$) in a given projected surface area of the molecular cloud if we make certain assumptions.  First, we assume a mean mass per YSO of 0.5 \msun, as suggested by \citet{2015ApJ...806..231H}, for the Class I and II stars  and for the ``indefinite" category which is expected to contain mainly those categories as well.  This value for the mean YSO mass should be correct to better than a factor of 2 for the cataloged objects.  Second, we must assume a mean age for all the YSOs in the sample.  We again follow \citet{2015ApJ...806..231H} who find that the duration of the Class I phase is 0.5 My, and that of the Class II phase is 2 My.  Our samples of YSOs have roughly equal mixes of the two classes, as well as a significant fraction in the ``indefinite" category, so we adopt a mean age of 1 My.  Again, we believe this value should be correct to better than a factor of 2.  Thus, the estimated current rate of star formation will be for a time range integrated over the past 1 My, and the values on the vertical axes in Figure \ref{fig:11} can be converted to the recent star formation rate (in \msun~My$^{-1}$ pc$^{-2}$) simply by multiplying by a factor of 0.5 (i.e., subtracting 0.3 in the logarithm).  Finally, we assume that the YSOs included in the analysis have all formed very close to their currently observed locations, so that the derived gas column density describes the conditions in that part of the molecular cloud where the stars were formed.  Note that only regions with detected CO are included, so the YSOs located where molecular gas is absent are not part of the calculation.  The observed molecular voids that contain YSOs have presumably been cleared of gas by stellar feedback processes, so the initial conditions of star formation at those positions are unknowable.  If such processes have also influenced the observed molecular gas, then the star formation---gas density relation may also be affected, a possibility we consider in section 5.

\begin{figure}
\plotone{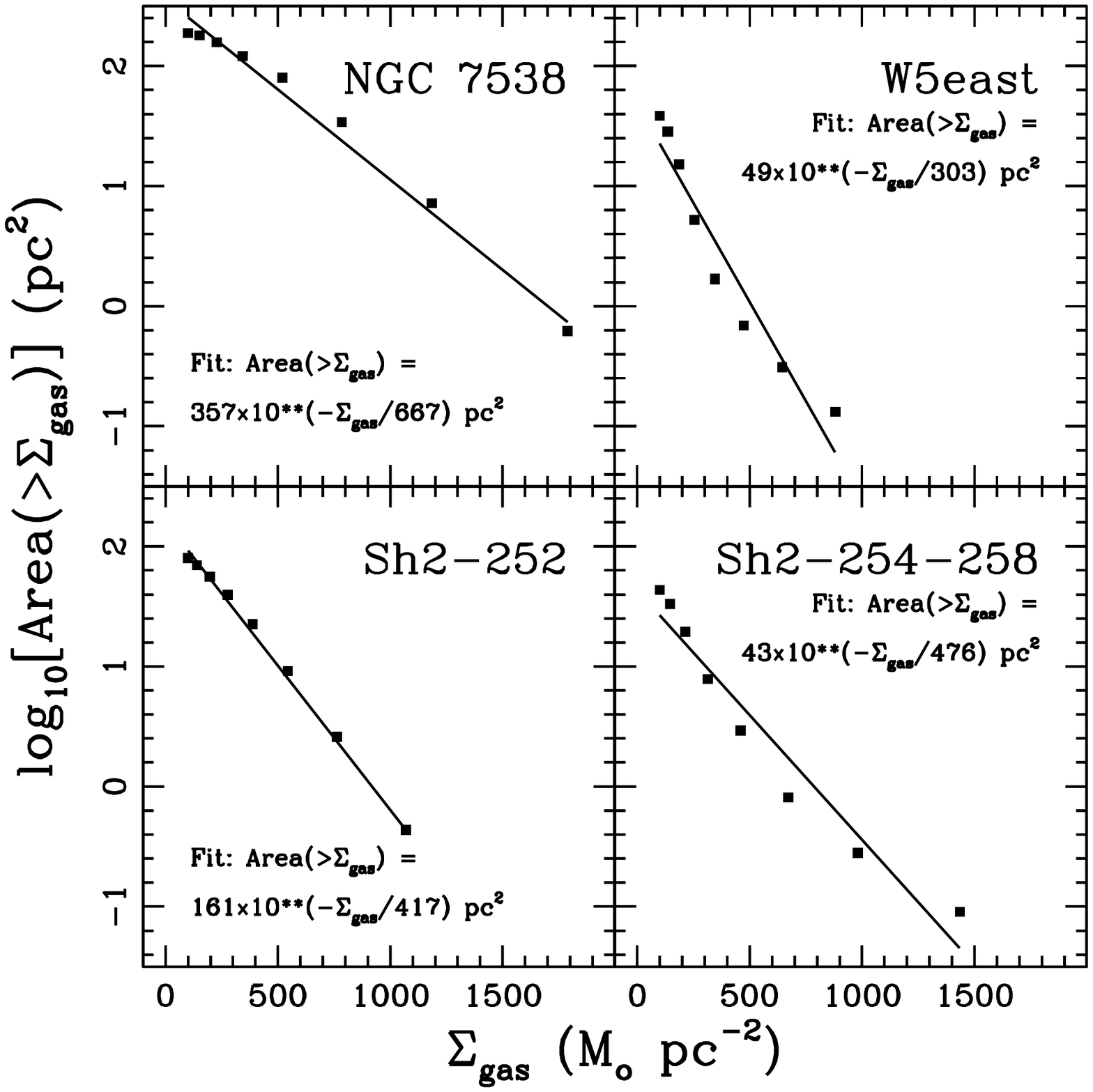}
\caption{The {\it structure functions}, defined as the area containing total molecular gas surface density greater than or equal to  $\Sigma_{gas}$ for each of the four mapped GMCs.  Exponential fits for surface density $\geq$100 \msun ~pc$^{-2}$ are shown.}\label{fig:12}
\end{figure}

The totality of star formation activity in a molecular cloud will depend critically on the distribution of gas density within the cloud.  We cannot measure the gas volume density directly, but we can examine the fraction of cloud projected area with column densities greater than or equal to $\Sigma_{gas}$, which we refer to as the {\it structure function}.  For Sh2-235, \citet{2016ApJS..226...13B} found that the projected area having $\Sigma_{gas}$~ above a given value was an exponential function of $\Sigma_{gas}$.  We have calculated the structure functions for the four GMCs in this study and show the results in Figure \ref{fig:12} as log-linear plots.   The points lie on nearly straight lines for 
$\Sigma_{gas} \geq 100$~\msun~ pc$^{-2}$, up to the maximum value.  We fitted exponential functions for each cloud, shown in Figure \ref{fig:12} together with the parameters of the best-fit exponential.  W5-east has a relatively steep slope compared to the other three GMCs, implying that it has a smaller fraction of cloud mass in the highest range of $\Sigma_{gas}$ compared to the others.

The aggregate star formation rate in these molecular clouds is the product of the power-law functions shown in Figure \ref{fig:11}, multiplied by the projected area within a given range of $\Sigma_{gas}$, and summed over the full range of gas column densities.   The projected area in a given range of $\Sigma_{gas}$ is the derivative of the structure function, which will therefore be another exponential.  The aggregate star formation rate will depend sensitively on the fraction of high column density gas in the cloud. If two clouds have similar masses but different structure functions, the cloud with the greater fractional area at high $\Sigma_{gas}$, i.e., a flatter structure function, should have the greater star formation rate per mass of molecular gas.  Therefore, from Fig. \ref{fig:12} we expect that NGC~7538 will exhibit more star formation than the other three clouds, per unit of cloud mass.

We note also that  \citet{2012ApJ...745..190L,2013ApJ...778..133L} have discussed the significant difference in star formation rates between the Orion Molecular Cloud and the so-called California Cloud, even though these two GMCs have similar total gas masses.  The difference, then, may be understood in terms of differences in the cloud density distributions or structure functions.  

\subsection{Properties of the molecular gas related to turbulence}

\begin{figure}
\plotone{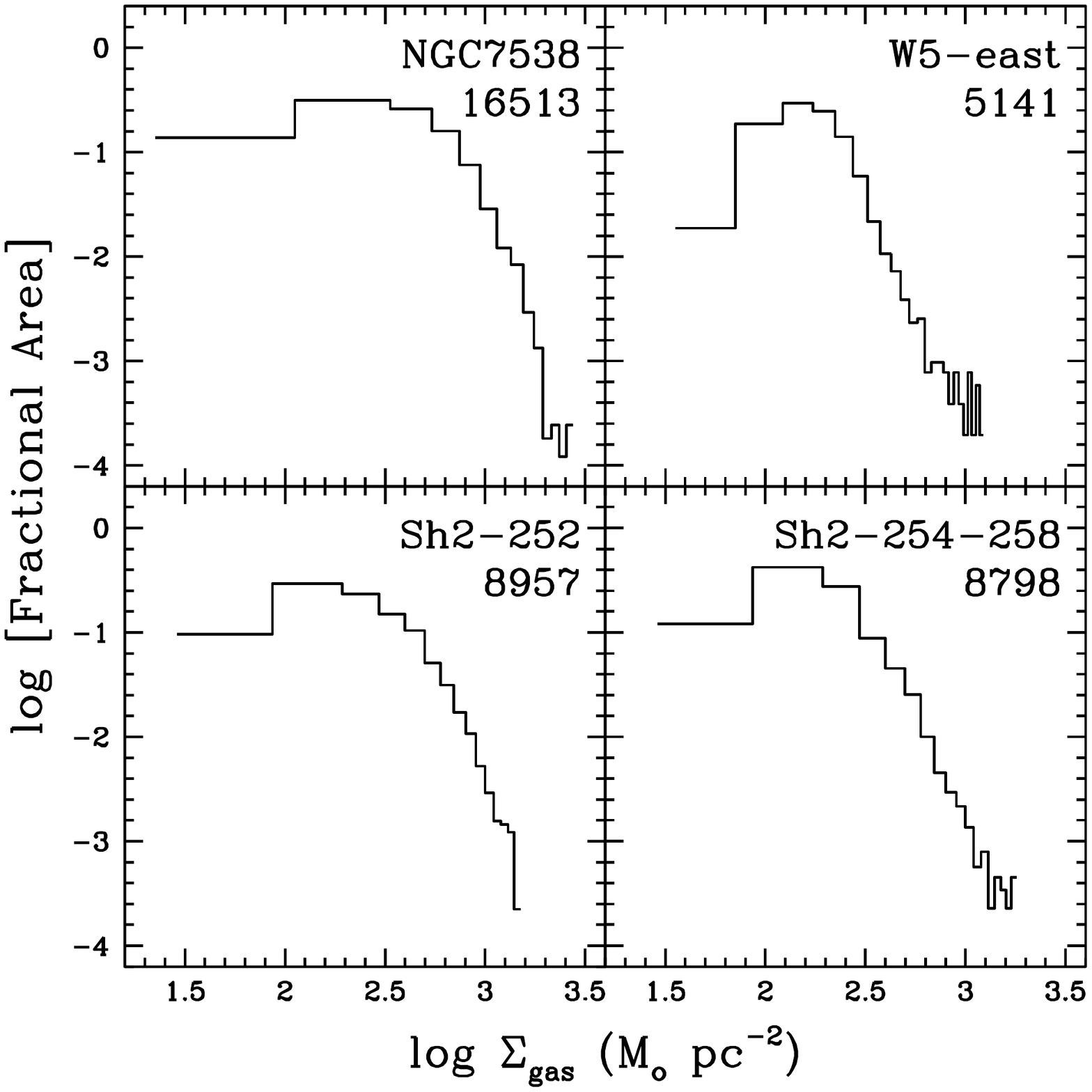}
\caption{Histograms of distribution of molecular gas column density in bins of 100 \msun ~pc$^{-2}$.  Numbers in upper right give the number of 10\arcsec ~pixels in each histogram.}

\label{fig:13}
\end{figure}

\begin{figure}
\plotone{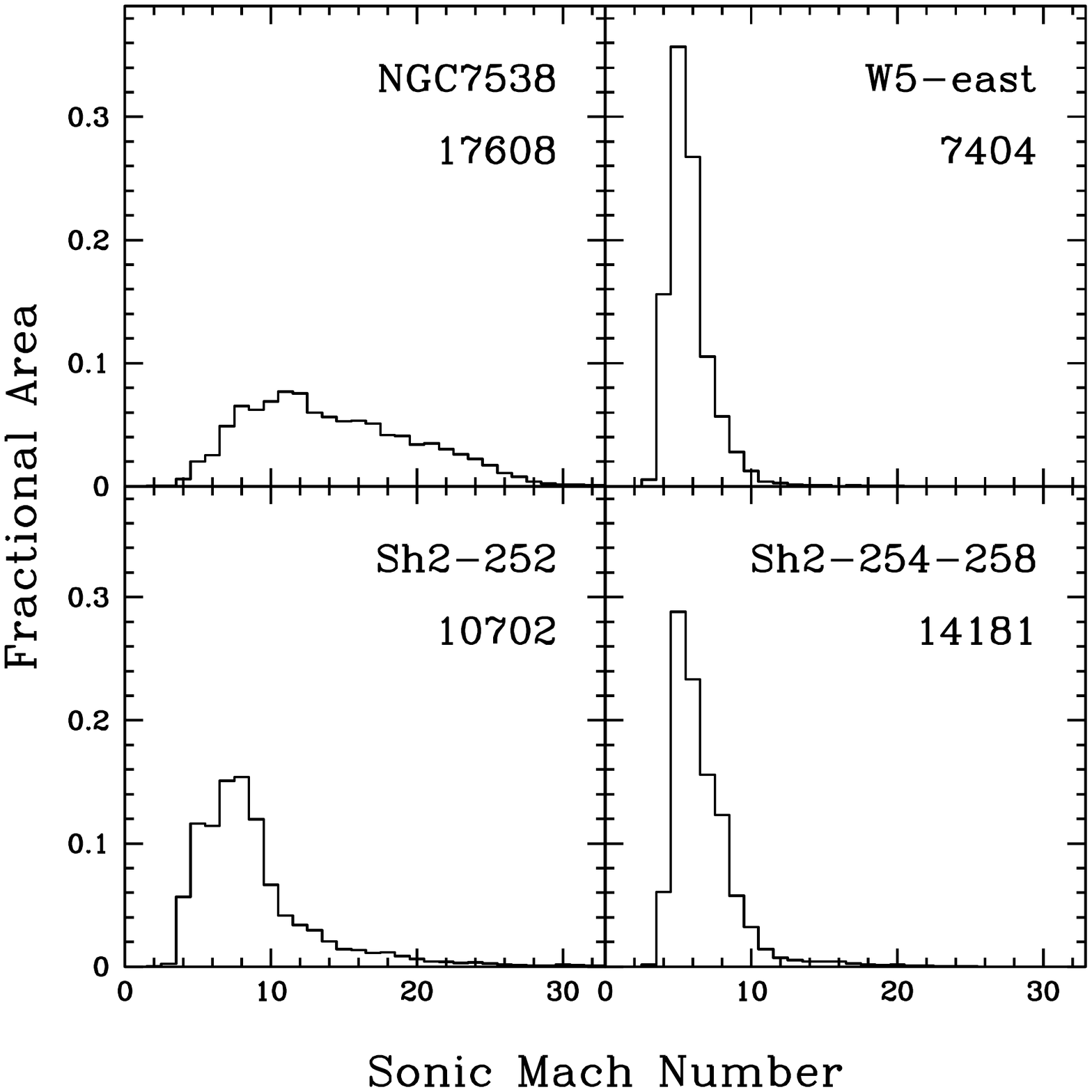}
\caption{Histograms of distribution of sonic Mach numbers for each GMC.   Numbers in upper right give the number of 10\arcsec ~pixels in each histogram. }\label{fig:14}
\end{figure}

In the Introduction we noted the importance of supersonic turbulence in currently favored theories of star formation.  Observational constraints on the properties of molecular clouds that are closely related to turbulence are therefore of great interest in the formulation of realistic numerical simulations.  Two properties that our molecular line maps allow us to calculate are the cloud column density PDFs (probability density functions) or $\Sigma_{gas}$-PDF, and the distribution of sonic Mach numbers.  In Figures \ref{fig:13} and \ref{fig:14} we show these for each of the four mapped GMCs.  As a proxy for the generally unmeasurable three-dimensional volume density distribution or $\rho$-PDF, the column density PDF has been shown to be related to the star formation activity in turbulent clouds, e.g., \citet{2014Sci...344..183K}.  These authors developed a formalism to derive a measure of the  $\rho$-PDF from the observed column density distribution such as those shown in Figure \ref{fig:13}. Previous studies have found a strong correlation between the rate of ongoing star formation and the flatness of the column density PDFs, e.g.,  \citet{2009A&A...508L..35K,2014Sci...344..183K, 2010ApJ...724..687L,2015A&A...577L...6S,2013ApJ...766L..17S}.  In a recent theoretical paper, \citet{2021MNRAS.507.4335K} show from numerical simulations that the $\rho$-PDF of GMCs with supersonic turbulence is initially a log-normal function but that a (broken) power-law tail develops at high volume densities as star formation progresses in the cloud.  Eventually a stable functional form is seen with a power-law dependence at the high end.  \citet{2021MNRAS.507.4335K} caution, however, that their numerical simulations do not include effects of stellar feedback or magnetic fields, which could further alter the shape of the $\rho$-PDF.  If the shape of the $\Sigma_{gas}$-PDF is related to the $\rho$-PDF as \citet{2014Sci...344..183K} argue, then our gas column density maps should reflect the effects of recent star formation.  

The $\Sigma_{gas}$-PDFs, shown in Figure \ref{fig:13}, do not extend to gas column densities low enough to reliably define the peak, so a log-normal function cannot with confidence be fitted to these data.  For $\Sigma_{gas} \geq 300$~\msun~ pc$^{-2}$ the distributions for W5-east and Sh2-254--258 are well-described by power-laws. Sh2-252 and NGC~7538 exhibit a steep high end that could be consistent with a log-normal function, but could also be well-described by a power-law.  All these GMCs are currently showing active star formation, so the extended power-law tail in the distribution function is consistent with results of the other studies noted above.  

The CO and \tco~line widths are generally much larger than would be observed if the only broadening were due to thermal motion in the molecular gas.  The excess width is presumably caused by bulk turbulent motions.  To calculate the sonic Mach number at each map position, we follow the approach used in \citet{2018ApJS..238...20B} where we assume that the excitation temperature of CO is equal to the gas kinetic temperature, $T_{kin}$ and the isothermal sound speed is given by $C_S = \sqrt{k T_{kin}/\mu}$, where $\mu = 2.33$~AMU including H$_2$ and He.  We assume the bulk gas motions are distributed isotropically and take the line-of-sight velocity dispersion from the second moment of the \tco~spectra, $\sigma_{\nu}$.  The \tco~spectra are less broadened by high optical depth compared to the CO spectra, so should be a more reliable measure of the actual gas motions. The 3-dimensional sonic Mach number is then given by 

$$M_s = \sqrt{3} \sigma_{\nu}/C_s .$$  

Figure \ref{fig:14} shows histograms of $M_s$ for the range 0 to 30 for each GMC.  All the pixels with CO emission have sonic Mach numbers well above 3, indicating that supersonic turbulence dominates the gas motions throughout the clouds.  For three GMCs, the distribution is sharply peaked at $M_s$ in the range 5 - 7, with the bulk of the cloud areas between 4 and 10.  Sh2-252 has a tail extending to $M_s = 20$, but most of the cloud has $M_s<10$.  NGC~7538 has a significantly broader distribution than the other three GMCs, with $M_s$ extending up to 30 and the maximum at 12.  NGC~7538 has several very luminous embedded O stars which may be generating stronger turbulent motions via stellar winds and supersonic expansion of the associated H~II region, compared to the other three.  The NGC~7538 cloud has the greatest molecular mass of our four targets, and as noted above, a shallower structure function than the others.    Together these properties imply a higher star formation rate which could be expected to produce a greater degree of supersonic turbulence compared to the other GMCs in our study.

\section{Comparison with the analysis of Pokhrel et al. (2021)}
\subsection{Calculation of cumulative star formation rates vs. mean gas column density}
In a recent paper, \citet{2021ApJ...912L..19P} have examined the molecular cloud column density---star formation rate relationship for a different set of clouds and an alternative approach to deriving the gas column density compared to the clouds and analysis we present in this paper.  We have applied their methodology to our data sets; here we compare our results with theirs.  In observational material, the main differences between \citet{2021ApJ...912L..19P} and the present work are:  (1) their clouds are relatively nearby, $d < 1$~kpc, while our four GMCs are between 1.6 and 2.7 kpc distant;  (2) their catalogs of YSOs were derived from a new analysis of 2MASS photometry together with archival data from the {\it Spitzer} satellite in the near- and mid-IR, while we use the YSO catalogs of \citet{2014MNRAS.439.3719C} which also relied on {\it Spitzer} data as well as deep JHK$_S$ photometry with large telescopes; (3) they include only the youngest YSOs, of Class 0/I with a mean age of 0.5 My, whereas we include Class 0/I and Class II objects and assume a mean age of 1 My;  (4) they inferred the column density distribution of molecular gas from {\it Herschel} far-IR images of thermal dust continuum, with an assumed dust absorption coefficient from standard models and a standard gas/dust ratio.  We derived the molecular gas column density from our observations of CO isotopologues and a \tco/H$_2$ abundance ratio, which we assume to be a function of the temperature-dependent CO depletion using the results of \citet{Lewis_2021}, which were based on {\it Herschel} photometry.  

Following the method of \citet{2021ApJ...912L..19P}, we calculate a  nominal mean gas volume density, $\av {\rho(\Sigma_{gas})}$, for a series of contours of constant $\Sigma_{gas}$ taken in the maps shown in Figures \ref{fig:7} - \ref{fig:10} for each of the four GMCs.  The mean gas volume density is defined as the total gas mass enclosed by a given contour at $\Sigma_{gas}$ divided by the volume of a sphere which has a projected area equal to the area enclosed by that contour.  From the mean gas volume density we calculate a nominal free-fall time with the usual expression, 

$$t_{ff}(\Sigma_{gas}) = \sqrt{3\pi/32 G \av{\rho(\Sigma_{gas})}} $$

\noindent for gravitational collapse of a uniform-density sphere with negligible pressure support.  \citet{2021ApJ...912L..19P} also define a mean gas {\it column} density as the total gas mass enclosed by the contour at a given value of $\Sigma_{gas}$ divided by the area within that contour.  We denote the mean gas column density by $\av{\Sigma_{gas}}$.  The free-fall time is therefore a function of the mean gas column density determined by  the enclosed mass and projected surface area at a given value of $\Sigma_{gas}$.   In our sample of GMCs, $t_{ff}$ varies by a full order of magnitude from the lowest calculated value at $\Sigma_{gas} = 100$~\msun~pc$^{-2}$ up to 2000~\msun~pc$^{-2}$, ranging typically from $\sim 1$~My down to $\sim 0.1$~My at the highest gas column density.  The projected area and enclosed mass vs. $\Sigma_{gas}$ are directly related to the structure function (Fig. \ref{fig:12}), which we find is reasonably well-described by an exponential function for $\Sigma_{gas} \geq 100$~\msun~pc$^{-2}$, for all four GMCs in our sample.  Likewise, a log-linear plot of $t_{ff}$ vs. $\av{\Sigma_{gas}}$ implies that the nominal free-fall time is an exponential function of the mean gas column density.  

We take a set of 8 contours equi-spaced in  $log (\Sigma_{gas})$ from 100 \msun~pc$^{-2}$ up to the maximum in each map, as in Section \ref{subsec:sfrsigma}.  For each contour of total gas column density we calculate a cumulative star formation rate from the total number of YSOs enclosed by each contour, divided by the enclosed area.  The mean mass per YSO is assumed to be 0.5 \msun~ and the average age to be 1 My, the same as in section \ref{subsec:sfrsigma}.  Note that this definition of the star formation rate differs from the one used in section \ref{subsec:sfrsigma}, which uses YSO average surface  densities restricted to binned ranges of $\Sigma_{gas}$.  The definition of the star formation rate in \citet{2021ApJ...912L..19P} sums over all the YSOs enclosed by the specified  $\Sigma_{gas}$~ contour up to the maximum $\Sigma_{gas}$~in the map; we refer to this as the {\it cumulative} star formation rate.

\begin{figure}

\plotone{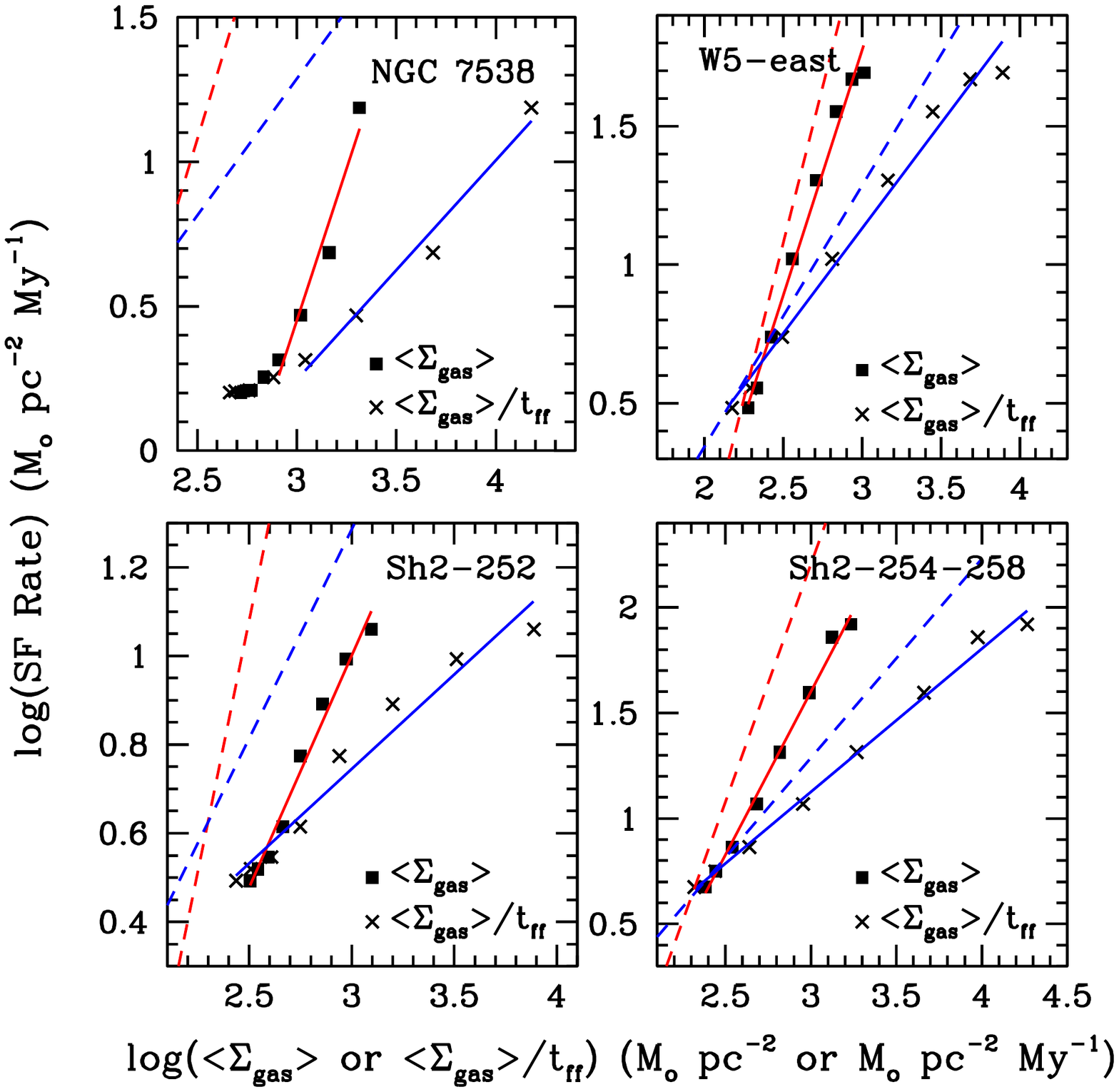}
\caption{Logarithmic plots of cumulative star formation rate vs. (1) mean gas column density,  $\av{\Sigma_{gas}}$ \it{(filled squares)} \rm or (2) mean gas column density divided by free-fall time, $\av{\Sigma_{gas}}$/$t_{ff}$ ($\times$'s), \rm following the method of \citet{2021ApJ...912L..19P}.  Solid lines are linear fits to the data points (see Table \ref{tab:fit}).  For NGC~7538, only the 4 highest points are used in the fits.  Dashed lines show the mean relations reported by Pokhrel et al. for their sample of molecular clouds.  Red lines represent log($\av{\Sigma_{gas}}$) and blue lines represent log($\av{\Sigma_{gas}}/t_{ff}$).  Note that as defined here, $t_{ff}$ is a function of $\av{\Sigma_{gas}}$ which, we find, is well-described as an exponential.}\label{fig:15}
\end{figure}

We show in Figure \ref{fig:15} a log-log plot of the cumulative star formation rate vs. the mean gas column density $\av{\Sigma_{gas}}$, where each plotted pair of values is calculated for the series of outer contours described above.  The points are well-described by straight lines for W5east, Sh2-252, and Sh2-254-258, implying a power-law relation between the two quantities.  For NGC~7538, the 4 highest points appear to follow a linear relation, while the lowest 4 points turn over to a shallower slope.  We fitted the points with a linear regression, shown as solid lines in the Figure.   The coefficients of the fits are summarized in Table \ref{tab:fit}, as well as the average for the four GMCs and the corresponding fit from \citet{2021ApJ...912L..19P} for the average of their 12 clouds.  For comparison we also list the coefficients of the power-law fits as described in Section 4.1 and shown in Figure \ref{fig:11}.  

 \citet{2021ApJ...912L..19P} also examined the relationship between cumulative star formation rate and the mean gas column density normalized by the corresponding free-fall time, $t_{ff}$~(as they define it) and found a smaller scatter in the fitted power-law functions.  They argued that this reduction in scatter of the fits is consistent with theoretical expectations and points to a universal correlation between star formation rate and (mean) gas column density normalized by free-fall time for all star-forming molecular clouds, where these quantities are calculated per their methodology.  

In Figure \ref{fig:15} we also show the cumulative star formation rate vs. mean gas column density normalized by $t_{ff}$ on the abscissa; these points are also well-described by power-laws whose values are given in Table \ref{tab:fit} along with the average values from \citet{2021ApJ...912L..19P}.  The figure shows the average fitted functions from  \citet{2021ApJ...912L..19P} as dashed lines. As shown in Figure \ref{fig:15} and Table \ref{tab:fit}, our power-law fits are somewhat shallower than the mean from \citet{2021ApJ...912L..19P}.  Their mean power-law exponent for the plot against log($\av{\Sigma_{gas}}$) is 2.00 with a standard deviation of 0.27 compared with a mean value of  1.62 for our four GMCs \citep[though the 4 highest points for NGC~7538 show a slope comparable to the mean from][]{2021ApJ...912L..19P}.  The fitted functions for the plots against log($\av{\Sigma_{gas}} / t_{ff}$) have a similar trend.  \citet{2021ApJ...912L..19P} find a nearly linear relation with a mean power-law exponent of 0.94 and standard deviation of 0.11, while our average value is 0.66. 
The discrepancies between our derived mean relations for star formation rate vs. $\Sigma_{gas}$ or vs.  $\Sigma_{gas}/t_{ff}$ (shown in Figure \ref{fig:15} and Table \ref{tab:fit}), and the mean relations from \citet{2021ApJ...912L..19P} are therefore at the level of $\sim 1.5\sigma$ and $\sim 2.5\sigma$ respectively.

Our analysis uses observations of the J=2-1 transitions of CO and \tco~ which we assume are populated very close to LTE conditions.  We justify this assumption based on our previous multi-transition CO observations of similar GMCs--Sh2-235 \citep{2016ApJS..226...13B} and Cep B/C \citep{2018ApJS..238...20B}.  The CO rotational levels through J=2 should be very close to LTE for the bulk of the molecular gas.  
The J=2-1 transition of CO is relatively insensitive to the assumed gas temperature for LTE conditions.  The volume emissivity changes by at most about $\pm 25$\% for kinetic temperatures from 10 K to 60 K \citep[see e.g.,][]{10.1111/j.1365-2966.2011.19279.x}.  We therefore argue that our measurements of the CO {\it gas} column densities are robust for the bulk of the star-forming gas.  To account for depletion of CO by freeze-out onto dust grains, we adopt the temperature-dependent [\tco] gas-phase abundance of \citet{Lewis_2021} to obtain the total gas column density.  \citet{Lewis_2021} derived their total gas column density from multi-wavelength {\it Herschel} far-IR maps of thermal dust emission from molecular clouds.  

\citet{2021ApJ...912L..19P} also employ far-IR maps of thermal dust emission from their sample of molecular clouds made by the {\it Herschel} satellite, together with a single assumed dust opacity coefficient and dust temperatures derived from fits to the SEDs at each map position.   They also use low spatial resolution maps from the {\it Planck} satellite to correct for the loss of low spatial frequency information in the {\it Herschel} data.  
The inferred dust column density is sensitive to the assumed dust temperature, which may vary with position along the line of sight within the cloud.   \citet{2021ApJ...912L..19P} assume a single temperature at each map pixel in the SED fits, a potential source of systematic bias if there are significant temperature gradients along the line of sight.  

To calculate the YSO surface density for our GMCs, we took all the stars in the \citet{2014MNRAS.439.3719C} catalogs, combining both Class I and II as well as the ``indefinite" category which is a mix of both classes.  \citet{2021ApJ...912L..19P} used only the class I objects in their new revision of the 2MASS-based catalogs of YSOs.  If the class II stars have diffused away from their birthplace over the course of early evolution from class I to II, we would expect that the older of the class II YSOs will show a more extended spatial distribution, provided all star formation has occurred in the same physical location.  Figures \ref{fig:3} - \ref{fig:6}  do in fact show that the class II stars have a more extended distribution than class I; the stars in the ``indefinite" catalog appear to be distributed more widely as well, comparable to class II.  Diffusion from their place of origin could tend to flatten the $log(\Sigma_{gas})$  -  $log(star~formation~rate)$ relation where we include the class II objects.  
To examine this possibility, we have also made a separate analysis for each of the 3 categories cataloged by \citet{2014MNRAS.439.3719C}.  
We find no evidence in these data, however,  that the use of combined class I and II YSOs in calculating the stellar surface density results in substantially flatter power-law dependence on gas column density compared to using only {\it bona fide} class I YSOs.  It appears, then, that this difference is not the explanation for the  shallower power-laws in our analysis compared to that of \citet{2021ApJ...912L..19P}.  We also note that in a previous paper, \citet{2020ApJ...896...60P} using both class I and II YSOs, obtained similar power-law fits as \citet{2021ApJ...912L..19P} did using only class I.
\subsection{Effects of stellar feedback}
Besides possible observational biases, the differences in power-law relations may result from real astrophysical differences between our sample of GMCs and that of \citet{2021ApJ...912L..19P}.  The four GMCs in our study are all actively forming massive stars and show clear evidence in their morphologies of stellar feedback mechanisms at work, principally the effect of expanding H~II regions acting on the surrounding molecular gas.  Of the 12 clouds in the sample of \citet{2021ApJ...912L..19P} at least half are relatively nearby regions forming mainly lower mass stars, not OB stars.  A few in their sample (Orion A, Sh2-140, Cep OB3, Cygnus X) are forming massive stars.  Nevertheless, they find that all have similar power-law relations between the mean gas column density normalized by free-fall time, $\av{\Sigma_{gas}/t_{ff}}$, and the mean YSO surface density using their cumulative definition for the surface densities and free-fall times.  

It would be instructive, then, to compare structure functions for the thermal dust continuum maps used by \citet{2021ApJ...912L..19P} with those shown in Figure \ref{fig:12} to see if exponential relations also apply to their dust-derived column density maps.  Our derivation of the  differential gas column density---star formation rate relation uses an alternative approach to the cumulative relations calculated by \citet{2021ApJ...912L..19P}, which integrate over a range of $\Sigma_{gas}$ values.  A calculation using the data from \citet{2021ApJ...912L..19P} would be interesting to compare directly with our results as shown in Figure \ref{fig:11}.  
The cumulative approach used by \citet{2021ApJ...912L..19P} assumes the simplest possible description of star-forming clouds, i.e., the uniform sphere, to derive the mean gas volume  density $\av{\rho_{gas}}$ and from that the free-fall time.  (\cite{2021MNRAS.502.5997H} have recently discussed the systematic errors  inherent in the assumption of uniform spherical clouds in deriving the free-fall time and from that the star formation efficiency.)  The density structure of GMCs is obviously far from spherical or uniform, so we argue that a more intuitive approach to determining the star formation rate---gas column density relation is the one we use to obtain the plots in Figure \ref{fig:11}, where column density is divided into discrete bins and YSO surface density is derived for each $\Sigma_{gas}$ bin.  These empirically derived relations can be compared directly with the predictions of numerical simulations, e.g. as shown in Figure 12 of \citet{2012ApJ...761..156F}.  

Despite these reservations, we find that the power-law fits for the binned ranges of $\Sigma_{gas}$ are very close to the values for fits to our data using the cumulative method of \citet{2021ApJ...912L..19P}, as shown in Table \ref{tab:fit}.  
The mean coefficients for our fits and those of \citet{2021ApJ...912L..19P} differ at the $1.5\sigma - 2.5\sigma$ level, with values for our sample generally lower.  NGC~7538 has coefficients close to the means of \citet{2021ApJ...912L..19P} but with large uncertainties and only over the highest range of $\Sigma_{gas}$.  We suggest that these differences may reflect the more advanced stage of star formation which our sample of GMCs represent compared to the clouds in the \citet{2021ApJ...912L..19P} analysis.  The morphologies of our four GMCs show clear signs of stellar winds or H~II region expansion having cleared out parts of the molecular gas, leaving stellar clusters behind.  The remaining molecular gas is evidently still forming stars, since we see clusters or isolated YSOs within these regions.  The molecular gas may be heated and compressed by the stellar radiation and/or H~II region expansion, so that further star formation in the gas is affected.  A good example is the H~II region group Sh2-254--258 (Fig. \ref{fig:9}), where a peak in both $\Sigma_{gas}$ and $T_{ex}$ lies directly between two young, nearly spherical H~II regions.  A tight cluster of YSOs coincides with the peak in the molecular gas.

Our derived star formation --- gas column density relations are not drastically different from those found by \citet{2021ApJ...912L..19P} except possibly for NGC~7538, where the relation at the high column density end has a slope close to their average, but for $\Sigma_{gas} < 500 $~\msun pc$^{-2}$ is nearly constant.  It would be of interest to compare the results of numerical simulations which incorporate stellar feedback, at different stages in the evolution of the cloud, with these observational results.  How do H~II region expansion or stellar wind input alter the conditions for star formation in the affected molecular gas?  Can simulations reproduce the relations observed in real GMCs?  How do these change as conditions evolve over time?  

A significant limitation on our analysis results from the need to examine GMCs  which are relatively distant compared to the 12 clouds in the \citet{2021ApJ...912L..19P} sample.  The bulk of star formation in the Galaxy occurs in GMCs, which mostly lie well beyond 1 kpc from the Sun.  Accurate photometry in the near-  and mid-IR is crucial to identifying and classifying YSOs, but for distances $> 1$ kpc, limitations in sensitivity and resolution hamper the available catalogs of YSOs, which have typically depended on observations with the {\it Spitzer} IRAC and MIPS-24$\mu$m instruments.  With the advent of {\it JWST}, we may expect significant improvements in our ability to identify and classify YSOs down to fainter luminosities in a larger set of GMCs.  It will also be important to examine massive GMCs in which the star formation process is still at an early stage (compared to the clouds in our sample), for example the many infrared dark clouds (IRDCs) which have now been identified and which have not yet formed OB stars leading to cloud disruption by stellar feedback.

\section{Summary}
We have presented fully sampled on-the-fly maps of 4 galactic GMCs (NGC~7538, Sh2-252, Sh2-254--258, and W5~east) in the J=2-1 emission of both CO and \tco.  After modest smoothing, the spatial resolution is 38\arcsec~ (FWHM) and the velocity resolution is 1.3 \kms, yielding average rms noise of $\sim 0.1$~K brightness temperature in one velocity channel.  We use an LTE analysis based on the CO peak brightness temperature and the \tco~ integrated intensity to derive maps of the CO excitation temperature and gas phase \tco column density.  To account for depletion of CO onto dust grain surfaces, we use the results of \citet{Lewis_2021} to derive the distribution of total molecular gas column density $\Sigma_{gas}$.  The surface density of YSOs of classes I and II is obtained from the YSO catalogs of \citet{2014MNRAS.439.3719C}. Our main conclusions are:

(1)  For all 4 GMCs, the mean YSO surface density within a series of binned gas column densities exhibits a power-law dependence on $\Sigma_{gas}$, with exponents in the range 0.9 to 1.9.  

(2)  The projected surface area having column density $\geq \Sigma_{gas}$, or {\it structure function}, is well-represented by a negative exponential function of $\Sigma_{gas}$ for column densities $\geq 100$~\msun ~pc$^{-2}$ up to the maximum in the map.  

(3)  Histograms of  $\Sigma_{gas}$ show power-law tails extending to high column densities, consistent with theoretical predictions for GMCs exhibiting current active star formation.

(4)  Histograms of the sonic Mach number are sharply peaked at $M_{sonic} \approx 5-8$ for 3 GMCs;  the fourth (NGC~7538) has a broad distribution that may result from effects of turbulent energy input from several massive stars recently formed.

(5)  We also analyze our data following the methodology of \citet{2021ApJ...912L..19P} and find that our sample of GMCs shows power-law relations that are somewhat shallower than found by \citet{2021ApJ...912L..19P} for the star formation rate vs. $\av{\Sigma_{gas}}$ and vs. $\av{\Sigma_{gas}}/t_{ff}$.  The average relations agree at the $\sim 1.5\sigma - 2.5\sigma$ level. 

(6)  We discuss possible effects of stellar feedback on the star formation -- gas density relations for our more evolved sample of GMCs compared with the clouds studied by \citet{2021ApJ...912L..19P}.

\acknowledgements

We thank Charlie Lada of the Harvard-Smithsonian Center for Astrophysics for fruitful discussions.

The Heinrich Hertz Submillimeter Telescope is operated by the Arizona Radio Observatory, which is part of Steward Observatory at The University of Arizona.  

The research presented in this paper has used data from the Canadian Galactic Plane Survey, a Canadian project with international partners, supported by the Natural Sciences and Engineering Research Council. 

This work has made use of data from the European Space Agency (ESA) mission
{\it Gaia} (\url{https://www.cosmos.esa.int/gaia}), processed by the {\it Gaia}
Data Processing and Analysis Consortium (DPAC,
\url{https://www.cosmos.esa.int/web/gaia/dpac/consortium}). Funding for the DPAC
has been provided by national institutions, in particular the institutions
participating in the {\it Gaia} Multilateral Agreement.

\bibliography{main}
\bibliographystyle{aasjournal}

\begin{deluxetable*}{lllcccc}
\tablecaption{Parameters of observed GMC fields \label{tab:gmc}}
\tablehead{
\colhead{Name} & 
\colhead{R.A.\tablenotemark{a}} & 
\colhead{Decl.\tablenotemark{a}} & 
\colhead{Distance} & 
\colhead{Field size} & 
\colhead{Field size} & 
\colhead{Resolution}\\
\colhead{} &
\colhead{(J2000)} & 
\colhead{(J2000)} &
\colhead{(kpc)} &
\colhead{(arcmin)} &
\colhead{(pc)} &
\colhead{(pc)}
}
\startdata
W5-east&	03$^{\rm h}$02$^{\rm m}$00$^{\rm s}$&	60\arcdeg32\arcmin00\arcsec& 2.0 & $30.2 \times 30.2$& $17.6 \times 17.6$& 0.37\\
Sh2-252&	06$^{\rm h}$09$^{\rm m}$20$^{\rm s}$&	20\arcdeg36\arcmin00\arcsec& 2.1& $29.0 \times 20.2$ & $17.7 \times 12.3$& 0.39\\
Sh2-254--258&	06$^{\rm h}$13$^{\rm m}$10$^{\rm s}$&	17\arcdeg58\arcmin00\arcsec&1.6&$29.5 \times 20.2$ &$13.7 \times 9.4$& 0.29\\
NGC~7538&	23$^{\rm h}$14$^{\rm m}$40$^{\rm s}$&	61\arcdeg28\arcmin00\arcsec&2.7&$29.5 \times 20.2$ & $23.2 \times 15.8$& 0.50\\
    \enddata
\tablenotetext{a}{center of mapped field}
\end{deluxetable*}

\begin{deluxetable*}{lcccc}
\tablecaption{Derived properties of observed GMC fields \label{tab:derive}}
\tablehead{
\colhead{Name} &
\colhead{Field area} &
\colhead{Unblanked} &
\colhead{Total gas mass\tablenotemark{a}} &
\colhead{Mean gas column}\\
\colhead{} &
\colhead{(pc$^2$)} & 
\colhead{fraction} &
\colhead{(\msun)} &
\colhead{density\tablenotemark{a} (\msun ~pc$^{-2}$)}
}
\startdata
W5-east &	308 &	0.16&	  8,100&	 167\\
Sh2-252 &	218 &	0.43& 26,900 &	 290\\
Sh2-254--258 &	129	&	0.41& 11,900&	 224\\
NGC~7538 &	367 &	0.77&	138,000&	487\\
    \enddata
\tablenotetext{a}{includes only unblanked pixels}
\end{deluxetable*}

\begin{deluxetable*}{lcccccccc}
\tablecaption{Fits\tablenotemark{a} to plots of log($\Sigma_{\rm gas}$),  log($\av{\Sigma_{\rm gas}}$), or log($\av{\Sigma_{\rm gas}}$/t$_{ff}$)  vs.  log(SF~rate)\tablenotemark{b} \label{tab:fit}}
\tablehead{
\colhead{GMC Name} &
\multicolumn{2}{c}{Range of $\Sigma_{\rm gas}$\tablenotemark{c}} &
\multicolumn{2}{c}{Fit to $\Sigma_{\rm gas}$\tablenotemark{d}} & 
\multicolumn{2}{c}{Fit to $\av{\Sigma_{\rm gas}}$\tablenotemark{e}} &
\multicolumn{2}{c}{Fit to $\av{\Sigma_{\rm gas}}$/t$_{\rm ff}$\tablenotemark{e}} \\
\colhead{} &
\colhead{min} &
\colhead{max} &
\colhead{slope} &
\colhead{intercept} &
\colhead{slope} & 
\colhead{intercept}&
\colhead{slope} & 
\colhead{intercept}}
\startdata
W5-east&	100&	1200& 1.89(.14)& -3.57(.36)&	1.77(.08)&	-3.24(.21)&	0.76(.04)&	-0.84(.12)\\
S252&		100&	1500& 0.88(.16) & -1.37(.41) & 1.05(.06)&	-1.84(.17)&		0.43(.03)&	-0.23(.10)\\
S254--258&	100&	2100& 1.51(.09) & -2.65(.25)  &	1.55(.05)&	-2.74(.13)&		0.68(.02)&	-0.61(.07)\\
NGC~7538\tablenotemark{f}&	520&	2700& 1.95(.38) & -5.08(1.17) &	2.11(.33)&	-5.58(1.03)&	0.76(.08)&	-1.74(.29)\\
\\
Mean$\pm$rms& - & - & 1.56$\pm$0.49 &  &			1.62$\pm$0.44 &	-3.35&		0.66$\pm$0.16&	-0.86\\
\\
Pokhrel et al.\tablenotemark{g}&  &  &  &  &\\
mean values& - & - & - & - & 2.00 & -4.11 &		0.94 & -1.53\\
    \enddata
\tablenotetext{a} {Values in parentheses are 1-$\sigma$ uncertainties.}
\tablenotetext{b}{The star formation (SF) rate assumes (1) the mean mass per YSO is 0.5 \msun~ and (2) the mean age of all the YSOs in the \citet{2014MNRAS.439.3719C} catalog is 1 Myr.  The free-fall time, t$_{\rm ff}$, is calculated following the method of \citet{2021ApJ...912L..19P}. }
\tablenotetext{c}{in \msun~ pc$^{-2}$}
\tablenotetext{d}{See Figure \ref{fig:11}.  }
\tablenotetext{e}{See Figure \ref{fig:15}.}
\tablenotetext{f}{Fit is only to highest 4 bins.}
\tablenotetext{g}{\citet{2021ApJ...912L..19P}}
\end{deluxetable*}

\clearpage

\clearpage

\end{document}